\documentclass[%
aip,
amsmath,amssymb,
reprint, 
onecolumn,
nofootinbib
]{revtex4-1}

\usepackage{graphicx} 
\usepackage[framemethod=TikZ]{mdframed}
\usepackage{tikz}
\usepackage[x11names]{xcolor}
\usepackage[most]{tcolorbox}

\usepackage{mathptmx}
\usepackage{mathtools}
\usepackage{amsthm}
\usepackage{thmtools}
\usepackage{amsmath}
\usepackage{amssymb}
\usepackage{cancel}
\usepackage{braket}
\usepackage{mathrsfs}
\usepackage{slashed}
\usepackage{comment}
\usepackage{bm}
\usepackage{etoolbox}
\usepackage{dsfont} 

\usepackage[T1]{fontenc}

\usepackage[colorlinks=true, linkcolor=blue, citecolor=blue, filecolor=blue, urlcolor=black]{hyperref} 
\usepackage{caption}
\usepackage{cleveref}

\makeatletter
\def\@email#1#2{%
 \endgroup
 \patchcmd{\titleblock@produce}
  {\frontmatter@RRAPformat}
  {\frontmatter@RRAPformat{\produce@RRAP{*#1\href{mailto:#2}{#2}}}\frontmatter@RRAPformat}
  {}{}
}
\makeatother
\begin{document}

\preprint{AIP/123-QED}

\title[Integrable Non-Holonomic Constraints and Gauge Fixing in Classical Field Theory]{Integrable Non-Holonomic Constraints and Gauge Fixing in Classical Field Theory}

\author{B. Bert}
\affiliation{\mbox{Department of Physics, University of Cape Town, Private Bag X3, Rondebosch 7701, South Africa}}
\author{W.A. Horowitz}
\affiliation{\mbox{Department of Physics, University of Cape Town, Private Bag X3, Rondebosch 7701, South Africa}}
\affiliation{\mbox{Department of Physics, New Mexico State University, Las Cruces, New Mexico 88003, USA}}
\affiliation{Theoretical Sciences Visiting Program, Okinawa Institute of Science and Technology Graduate University, Onna, 904-0495, Japan}

\email{wahorowitz@uct.ac.za}
\email{brtben004@myuct.ac.za}

\date{\today}

\begin{abstract}
We re-examine the derivation of the equations of motion from an action principle for classical field theories with non-holonomic constraints, \textit{i.e.}, constraints involving derivatives of the fields. We find that the usual method for gauge fixing in classical and quantum field theories is highly non-trivial for non-holonomic gauge constraints, such as the Coulomb and Lorenz gauges. The subtlety appears at the use of the so-called transposition rule, $\delta(\partial_\nu A^\mu)=\partial_\nu(\delta A^\mu)$, which has been shown not to hold for general non-holonomic constraints in the point-particle context. We provide a sufficient definition of integrable non-holonomic constraints in classical field theory that allows us to prove that the transposition rule holds for all theories with these constraints; we are then able to recover the usual treatment of gauge fixing for gauges of this type.
\end{abstract}

\maketitle

\makeatletter
\renewcommand*\l@subsection[2]{}
\renewcommand*\l@subsubsection[2]{}
\makeatother
\tableofcontents

\section{Introduction}
\label{Introduction}

The dynamics of classical field theories are often determined by the Euler-Lagrange equations of motion, derived from Hamilton's principle\cite{Hamilton1834,Galley:2012hx,rothkopf2024unifyingactionprincipleclassical}. Crucial to the derivation of the Euler-Lagrange equations from Hamilton's principle is the \textit{transposition rule}, which in point particle mechanics takes the form
\begin{equation}
    \delta\bigg(\frac{d q^i}{d t}\bigg) = \frac{d}{d t} \bigg(\delta q^i\bigg),
\end{equation}
where $q^i$ is the $i^{th}$ generalized coordinate of the particle. Flannery \cite{Flannery:2011} showed, in the context of point particle mechanics, that the transposition rule does not hold when general non-holonomic constraints $g_j(q^i,\dot q^i,t)=0$ are present, but that the transposition rule does hold when the non-holonomic constraints are integrable, \textit{i.e.}, when the non-holonomic constraints can be expressed as the total time derivative of some function $f_j(q^i,t)$ that is only dependent on the generalized coordinates $q^i$ and time $t$
\begin{equation}
    g_j(q^i,\dot q^i,t)=\frac{d f_j(q^i,t)}{dt}=0.    
\end{equation}

Many classical and quantum field theories exhibit gauge symmetry, a reflection of a redundancy in the fundamental degrees of freedom of the theory.
For example, the Standard Model of particle physics is constructed using classical gauge-invariant field theories\cite{Glashow1961,Higgs1964,Englert1964,Guralnik1964,GellMann1964,Weinberg1967,Salam1968,Gross1973,Politzer1973}. Many commonly used gauges, such as the Coulomb and Lorenz gauges, are specified by gauge conditions involving derivatives of the field components. These gauge conditions---which are ubiquitous in the literature
\cite{Dirac1964,yndurain1983qcd,Ryder_QFT,GaugeTheoriesOfTheStrongAndElectroweakInteraction,Henneaux:1992ig,Peskin:IntroQFT,Weinberg_QFT1,Weinberg_QFT2,Zee:2003mt,Srednicki:2007qs,Coleman:2018mew}---therefore impose non-holonomic constraints on the fields.

Traditional treatments of Hamilton's principle for classical field theories under non-holonomic constraints gloss over the subtleties of the functional analysis techniques being utilized; as such, the validity of the transposition rule in the presence of these non-holonomic constraints is often overlooked and simply assumed without justification
\cite{Dirac1964,yndurain1983qcd,Ryder_QFT,GaugeTheoriesOfTheStrongAndElectroweakInteraction,Henneaux:1992ig,Peskin:IntroQFT,Weinberg_QFT1,Weinberg_QFT2,Zee:2003mt,Srednicki:2007qs,Coleman:2018mew}. In this manuscript, we present a detailed study of functionals and functional derivatives allowing us to develop the necessary tools to carefully examine the validity of the transposition rule in classical field theories with non-holonomic constraints. We then apply these tools to carefully re-examine the derivation of the equations of motion from Hamilton's principle in the presence of non-holonomic constraints. The results of our careful analysis lead us to conclude that the assumption of the transposition rule is far from trivial in the context of classical field theory. Similar to point particle mechanics, the transposition rule
\begin{equation}
\label{TranspositionRuleEq}
    \delta(\partial_\nu A^\mu)=\partial_\nu(\delta A^\mu)
\end{equation}
is expected to not hold for field theories with general non-holonomic constraints.

One is then naturally led to ask: is there a concept of integrability for non-holonomic constraints in field theories, and does the transposition rule hold for field theories with these types of constraints? In this work, we provide a definition of integrable non-holonomic constraints in the context of field theory and show that this definition is sufficient for the transposition rule to hold under these types of constraints.  We show that the Coulomb and Lorenz gauges are particular examples of our definition of integrable, non-holonomic constraints, and thus one can justify the usual gauge fixing treatment of these gauges in classical field theory. In contrast to the Coulomb and Lorenz gauges, there exist gauges used in the literature that are non-holonomic and non-integrable, \textit{e.g.} the 't Hooft-Veltman gauge
\cite{Koplik:1978je,Mann:1984kam,Mckeon:1985vcr,Parthasarathy:1988zf,Gracey:2005vu,Gracey:2007rz,deGracia:2019qwj,Tran:2022fdb,Phan:2023xgi}. The 't Hooft-Veltman gauge---as per our definition---constitutes a non-integrable, non-holonomic constraint, and we therefore expect the standard application of Hamilton's principle to fail when fixing to this gauge. 

\section{Proof of the Transposition Rule for Integrable Non-Holonomic Constraints}
\label{ProofOfTranspositionRuleSection}
Suppose we are to impose a constraint $g$ on the spacetime derivatives of the field components $\partial_\nu A^\mu$ of a classical field theory in $d$ dimensions. Here and in what follows, we shall assume that all field components $A^\mu$ and their spacetime derivatives $\partial_\nu A^\mu$ belong to the function space of all infinitely differentiable functions on Minkowski spacetime, denoted here as $C^{\infty}(\mathbb{M})$. The spacetime manifold is taken to be
$
\mathbb{M}=[t_i,t_f]\times\mathbb{R}^{d-1}
$, 
where $t_i<t_f$ are real numbers and the manifold $\mathbb{M}$ is over a flat metric $\eta^{\mu\nu}=\text{diag}(1,-1,\ldots,-1)$. As is common in the physics literature, we will further require that the $A^\mu$ fields vanish at the spatial boundaries of $\mathbb{M}$, \textit{i.e.} $A^\mu(x)=0$ as $x^i\to \pm \infty ~\forall ~ i\in\{1,\ldots,d-1\}$; by imposing such a condition, we are excluding the possibility of plane wave solutions.

We shall further require that the constraint $g$ is satisfied for all spacetime points $x\in\mathbb{M}$, so that we may write
\begin{equation}
    \label{originalNonHolonomicConstraint}
    g[\partial_\nu A^\mu](x)=0, \quad \forall x\in\mathbb{M},
\end{equation}
where the square brackets indicate that $g$ depends functionally on derivatives of the field components $\partial_\nu A^\mu$, and the parentheses indicate dependence on the spacetime point $x$. The derivatives of the field components $\partial_\nu A^\mu$ are to be considered as independent from the field components $A^\mu$ as is usually done in the literature\cite{Dirac1964,yndurain1983qcd,Ryder_QFT,GaugeTheoriesOfTheStrongAndElectroweakInteraction,Henneaux:1992ig,Peskin:IntroQFT,Weinberg_QFT1,Weinberg_QFT2,Zee:2003mt,Srednicki:2007qs,Coleman:2018mew}.

Now consider the following vector functional $f^\nu=f^\nu[A^\mu](x)$, which is dependent on the field components $A^\mu$ but independent of the spacetime derivatives of the field components $\partial_\nu A^\mu$. Suppose that one of the field components, which we will denote as $A^{\mu_D}$, is specially singled out by $f^\nu$ in the sense that one of the spatial components of the vector functional $f^\nu$ can be written as\footnote{One could take $ f^{\mu_D}=r A^{\mu_D}$ for some $r>0$; in such a case, one could just rescale all field components by $1/r$ to recover \cref{specialCoordinate}.}
\begin{equation}
    \label{specialCoordinate}
    f^{\mu_D}=A^{\mu_D} \text{ for some } \mu_D\in\{1,\dots,d-1\},
\end{equation}
and that the remaining components of $f^\nu$ are independent of $A^{\mu_D}$, \textit{i.e.}
\begin{equation}
   f^{\nu}=f^{\nu}[A^\mu:\mu\neq\mu_D](x) \text{ for all } \nu\neq \mu_D.
\end{equation}
The field component $A^{\mu_D}$ will be referred to as the \emph{dependent} field component, and we will refer to the remaining field components $A^\mu$ with $\mu\neq\mu_D$ as the \emph{independent} field components. Lastly, we will require that $f^\nu$ with $\nu\neq\mu_D$ can be written as functional power-series of the independent field components $A^\mu$ with $\mu\neq\mu_D$; \textit{i.e.}
\begin{equation}
    \label{fNuAsPowerSeriesOfIndependentFieldComponents}
    f^\nu[A^\mu:\mu\neq\mu_D](x)
    =
    \sum_{I=0}^{\infty} 
    c^\nu_{i_0,\ldots,i_{d-1}}
    (A^0(x))^{i_0} \ldots(A^{\mu_D-1}(x))^{i_{\mu_D-1}}(A^{\mu_D+1}(x))^{i_{\mu_D+1}} \ldots(A^{d-1}(x))^{i_{d-1}},
\end{equation}
where the index set $I$ runs over all $d-1$ tuples of non-negative integers $\{i_0,\ldots,i_{\mu_D-1},i_{\mu_D+1}\ldots,i_{d-1}\}$, and the coefficients $c^\nu_{i_0,\ldots,i_{d-1}}$ are real numbers. 

Our proof of the transposition rule shall be presented for \textit{integrable non-holonomic} constraints, which we define to mean that the non-holonomic constraint $g$ can be expressed as the covariant divergence of a vector functional $f^\nu$ which satisfies the conditions mentioned above; \textit{i.e.}
\begin{equation}
    \label{integrabilityOfConstraint}
    0=g[\partial_\nu A^\mu](x) \equiv\sum_{\nu=0}^{d-1} \partial_\nu f^\nu[A^\mu](x), \quad \forall x\in\mathbb{M},
\end{equation} 
For the sake of simplicity, we shall restrict ourselves to the case where $g$ and $f^\nu$ are only implicitly dependent on the spacetime coordinates $x$ through their dependence on the derivatives of the field components $\partial_\nu A^\mu$ and the field components $A^\mu$, respectively. In this manuscript, we make use of explicit summation---rather than the Einstein summation convention---to prevent possible confusion later, as the summation will often be limited to run over a subset of the field components. We would like to further point out that the definition of integrability we have given in \cref{integrabilityOfConstraint} includes the restrictions we have imposed on the vector functional $f^\nu$; thus, the integrability condition in \cref{integrabilityOfConstraint} is a \emph{sufficient} condition for the transposition rule to hold, but it is not---as far as the authors can tell---a \emph{necessary} condition. Our definition of integrability will, however, be general enough to include the Coulomb and Lorenz gauges as particular examples (see \cref{ClassicalElectrodynamicsSection} for further details).

With these conventions in place, we have the following:
\begin{subequations}
    \begin{align}
        \partial_{\mu_D}f^{\mu_D}[A^{\mu_D}](x)
        &=
        -
        \sum_{\nu\neq\mu_D} \partial_\nu f^\nu [A^\mu:\mu\neq\mu_D](x)\\
        \label{explicitFromOfDependentFieldComponent1}
        \implies
        A^{\mu_D}[A^\mu:\mu\neq\mu_D](x)
        &=
        -\int_{-\infty}^{x^{\mu_D}} 
        d\tilde{x}^{\mu_D}~ 
        \sum_{\nu\neq\mu_D} \partial_\nu f^\nu [A^\mu:\mu\neq\mu_D](x^0,\ldots,\tilde{x}^{\mu_D},\ldots,x^{d-1}),
    \end{align}
\end{subequations}
where we have used the assumption that the field component $A^{\mu_D}$ vanishes as $x^{\mu_D}\to-\infty$. If we had allowed $\mu_D=0$ in \cref{specialCoordinate} then we would have required $A^{0}$ vanishes as $x^{0}\to-\infty$, which is generally an undesirable condition. In what follows, it will be useful to define the following functional
\begin{equation}
    \label{definitionOfBFunctional}
    B=B[\partial_\nu A^\mu: \nu,\mu\neq\mu_D](x)
    \equiv
    -\sum_{\nu\neq\mu_D} \partial_\nu f^\nu [A^\mu:\mu\neq\mu_D](x).
\end{equation}
\Cref{definitionOfBFunctional} will then allow us to rewrite \cref{explicitFromOfDependentFieldComponent1} as
\begin{equation}
    \label{explicitFromOfDependentFieldComponenInTermsOfBFunctional}
    A^{\mu_D}[\partial_\nu A^\mu: \nu,\mu\neq\mu_D](x)
    =
    \int_{-\infty}^{x^{\mu_D}} 
    d\tilde{x}^{\mu_D}~ 
    B[\partial_\nu A^\mu: \nu,\mu\neq\mu_D](x^0,\ldots,\tilde{x}^{\mu_D},\ldots,x^{d-1}).
\end{equation}

\subsection{Functional Calculus with Dependent Field Components}
\label{FunctionalFormalismSection}
As is evident from \cref{explicitFromOfDependentFieldComponenInTermsOfBFunctional}, the non-holonomic constraint $g$ resulted in the $A^{\mu_D}$ field component being functionally dependent on the derivatives of the independent field components $\partial_\nu A^{\mu}$ with $\nu,\mu\neq \mu_D$. We are ultimately interested in computing all the variations $\delta(\partial_\nu A^\mu)$ of the derivatives of the field components $A^\mu$; thus, we are tasked with defining the variation of the independent field components $A^{\mu}\in C^{\infty}(\mathbb{M})$ with $\mu\neq \mu_D$. Due to the functional dependence of $A^{\mu_D}$, we should expect that the variations $\delta A^{\mu_D}$ of the dependent field component will induce variations in the independent field components; likewise, we should expect the derivatives of the dependent field component $\partial_\nu A^{\mu_D}$ to induce derivatives in the independent field components $\partial_\nu A^{\mu}$. The purpose of this section is to make these expectations precise. In this section we will adapt the approaches taken in \cite{greiner1996field,friedlander1998distributions,engel2001dft,Cheney2001} to develop the functional framework necessary to prove the transposition rule.

We begin by defining a \textit{functional} $F$ as a map from the space of $n$-tuples of infinitely differentiable functions on $\mathbb{M}$, denoted as $(C^{\infty}(\mathbb{M}))^n$, to the real numbers $\mathbb{R}$, where $n\in\mathbb{N}$. We denote the space of all such functionals by $\mathscr{F}$; \textit{i.e.}
\begin{equation}
    \label{functionalDefinition}
    F\in\mathscr{F} \implies F[h_1(\cdot),\ldots,h_n(\cdot)]\in\mathbb{R}~~~ \forall~(h_1,\dots,h_n)\in (C^{\infty}(\mathbb{M}))^n.
\end{equation}
In \cref{functionalDefinition}, we use the notation $(\cdot)$ to indicate that each function $h_i$ requires an argument from $\mathbb{M}$; we will use the notation $(x)$ to indicate that we are evaluating the argument $(\cdot)$ at the specific point $x\in\mathbb{M}$. Now, to define a \emph{functional derivative}, we consider the following object
\begin{equation}
    \label{FunctionalAsAFunctionOfAlpha}
    F(\alpha|\gamma_i)\equiv F[h_1(\cdot),\ldots,h_i(\cdot)+\alpha \gamma_i(\cdot),\ldots,h_n(\cdot)],
\end{equation}
where $F(\alpha|\gamma_i)$ is to be understood as a function of the real,  spacetime independent parameter $\alpha$; furthermore, the $\gamma_i\in C^\infty_c(\mathbb{M})$ in \cref{FunctionalAsAFunctionOfAlpha} is a test function from the space of all infinitely differentiable functions with compact support that vanish at the temporal boundaries---\textit{i.e.} $\gamma_i(x)=0 ~\forall~ x\in\{t_i,t_f\}$. By ensuring that the test functions vanish at the temporal boundaries, we will be able to make use of integration by parts when applying our formalism to Hamilton's principle in \cref{EulerLagrangeEquationsSection}.

In what follows, we will assume that $F(\alpha|\gamma_i)$ is differentiable with respect to $\alpha$ around $\alpha=0$.

Next, we follow \cite{friedlander1998distributions} and define a \emph{distribution} as a linear functional $D$ which maps from the space of test functions $C^\infty_c(\mathbb{M})$ to the real numbers $\mathbb{R}$ such that for every compact set $\mathbb{K}\subset\mathbb{M}$, there is a real number $\tau\geq0$ and an integer $N\geq0$ such that
\begin{equation}
    \label{distributionDefinition}
    |D[\gamma(\cdot)]|
    \leq 
    \tau \sum_{|\kappa|\leq N} \sup_{\mathbb{K}} 
    \left|
        \frac{\partial^{|\kappa|} \gamma(\cdot)}{\partial x_0^{\kappa_0}\cdots\partial x_{d-1}^{\kappa_{d-1}}}
        \right|, 
    \quad \forall~\gamma\in C^\infty_c(\mathbb{K}),
\end{equation}
where $|\kappa|=\kappa_0+\ldots+\kappa_{d-1}$ is the magnitude of a multi-index of non-negative integers $\kappa=(\kappa_0,\ldots, \kappa_{d-1})$ and $d$ is the dimension of the manifold $\mathbb{M}$. We shall denote the space of all distributions on $\mathbb{M}$ as $\mathscr{D}'(\mathbb{M})$. The prime on $\mathscr{D}'(\mathbb{M})$ is standard notation used to indicate that the space of distributions is dual to the space of test functions. Note that the inequalities in \cref{distributionDefinition} are called semi-norm estimates, and their structure is motivated through topological considerations of the space of test functions $C^\infty_c(\mathbb{M})$; see \cite{friedlander1998distributions} for further details.

Now, suppose $F:(C^{\infty}(\mathbb{M}))^n\to\mathbb{R}$ is a functional and define $D_{G}F_i:C^\infty_c(\mathbb{M})\to\mathbb{R}$ as
\begin{equation}
    \label{GateauxDerivativeAsInnerProduct}
    D_{G}F_i[\gamma_i(\cdot)] 
    \equiv
    \int d^d x \frac{\partial F}{\partial h_i(x)} \gamma_i(x),
\end{equation}
where $\partial F/\partial h_i(x)$ is a function in $C^{\infty}(\mathbb{M})$. $D_{G}F_i$ in \cref{GateauxDerivativeAsInnerProduct} satisfies the conditions in \cref{distributionDefinition}\cite{friedlander1998distributions} and thus $D_{G}F_i$ is a distribution; furthermore, $D_{G}F_i$ is uniquely defined by the function $\partial F/\partial h_i(x)\in C^{\infty}(\mathbb{M})$ up to a set of measure zero\cite{friedlander1998distributions}. Note that the function $\partial F/\partial h_i(x)$ has no dependence on the test function $\gamma_i(x)$. Now, if the following limit holds for all $\gamma_i\in C^\infty_c(\mathbb{M})$
\begin{equation}
    \label{GateauxDerivativeDefinition}
    \lim_{\alpha\to0} 
    \frac{
        \big(F(\alpha|\gamma_i)-F(0|\gamma_i)\big)
        -
        \alpha D_{G}F_i[\gamma_i(\cdot)]}
        {\alpha}
    =
    0,
\end{equation}
then we call $D_{G}F_i$ the \emph{G\^ateaux derivative} \cite{Gateaux1913,Cheney2001,Dugger2014} of the functional $F$ with respect to the function $h_i$ in the direction of $\gamma_i$ and $\partial F/\partial h_i(x)$ is the \emph{functional derivative} of $F$ with respect to $h_i$ at the point $x\in\mathbb{M}$. In \cref{GateauxDerivativeAsInnerProduct}, we have left the bounds of integration implicit, but it is to be understood that the integration is over the entire manifold $\mathbb{M}$. 

By making use of \cref{GateauxDerivativeDefinition,GateauxDerivativeAsInnerProduct}, we shall often write the following
\begin{equation}
    \label{functionalDerivativeDefinition}
    \int d^d x \frac{\partial F}{\partial h_i(x)} \gamma_i(x) 
    = 
    \frac{\partial F[h_1(\cdot),\ldots,h_i(\cdot)+\alpha \gamma_i(\cdot),\ldots,h_n(\cdot)]}{\partial \alpha}\Bigg|_{\alpha=0}
    \equiv
    \lim_{\alpha\to0} 
    \frac{
        F(\alpha|\gamma_i)-F(0|\gamma_i)}
        {\alpha}.
\end{equation}
Taking the Dirac delta distribution $\delta^d(x)=\lim_{\beta \to 0}\delta_\beta^d(x)$ as the limit of a sequence of functions in $C^\infty_{c}(\mathbb{M})$, one may choose the arbitrary test function as $\gamma_i(x)=\delta_\beta^d(x-y)$ for some fixed but arbitrary $y\in\mathbb{M}$. \Cref{functionalDerivativeDefinition} then yields
\begin{equation}
    \label{gammaAsDeltaFunction}
    \frac{\partial F}{\partial h_i(y)} = \frac{\partial F[h_1(\cdot),\ldots,h_i(\cdot)+\alpha \delta^{d}(\cdot - y),\ldots,h_n(\cdot)]}{\partial \alpha}\Bigg|_{\alpha=0},
\end{equation}
where the limits involved in defining the delta distribution are implied but suppressed for notational simplicity. Notice the use of the notation ``$(\cdot-y)$'' which indicates that the argument of the function in question (in this case the delta function) is to be shifted by $y$. In what follows, we shall assume that for all functionals that we consider the limit in \cref{GateauxDerivativeDefinition} exists for all test functions in $C^\infty_{c}(\mathbb{M})$. 

Now, we define the variation $\delta F$ of the functional $F$ as:
\begin{subequations}
    \label{variationDefinition}
    \begin{align}
        \delta F 
        &\equiv \sum_{i=1}^{n} 
        \bigg(F(\alpha|\gamma_i)-F(0|\gamma_i)\bigg)\\
        \label{variationDefinitionExpandedITOderivative}
        &=
        \sum_{i=1}^{n} 
        \left(
        \frac{\partial F[h_1(\cdot),\ldots,h_i(\cdot)+\alpha \gamma_i(\cdot),\ldots,h_n(\cdot)]}{\partial \alpha}\Bigg|_{\alpha=0}
        \right)
        \alpha
        +\mathcal{O}(\alpha^2)\\
        &=
        \sum_{i=1}^{n} \int d^d x \frac{\partial F}{\partial h_i(x)} \gamma_i(x) \alpha
        +\mathcal{O}(\alpha^2).
    \end{align}
\end{subequations}
In this work, we shall concern ourselves with infinitesimal variations, and thus we shall drop all terms of order $\mathcal{O}(\alpha^2)$ and higher. Note that the variation $\delta F$ defined in \cref{variationDefinition} is dependent on the choice of $\{\gamma_i:i=1,\ldots,n\}$, and thus the variation $\delta F$ is to be thought of as a variation in the directions determined by $\{\gamma_i:i=1,\ldots,n\}$. To be more precise, one should write $\delta F$ as $\delta_{\gamma} F$ to indicate the dependence on the choice of $\{\gamma_i:i=1,\ldots,n\}$, but this notation will prove to be cumbersome.

Notice that \cref{explicitFromOfDependentFieldComponenInTermsOfBFunctional} implies that the dependent field component $A^{\mu_D}$ is functionally dependent on the derivatives of the independent field components $\partial_\nu A^{\mu}$, with $\nu,\mu\neq \mu_D$; however, $A^{\mu_D}$ is also a function of the spacetime coordinates $x\in\mathbb{M}$. Therefore, $A^{\mu_D}$ is a mapping from $(C^{\infty}(\mathbb{M}))^{d-1}$ to $C^{\infty}(\mathbb{M})$, rather than a functional as defined in \cref{functionalDefinition}. We are thus required to consider more general mappings than functionals; we shall refer to these more general mappings as \textit{function-valued functionals}. A function-valued functional $G$ is a map from $(C^{\infty}(\mathbb{M}))^m$ to $C^{\infty}(\mathbb{M})$ where $m\in\mathbb{N}$, and we shall denote the space of all such function-valued functionals as $\mathscr{G}$; \textit{i.e.}
\begin{equation}
    \label{functionalValuedFunctionalDefinition}
    G\in\mathscr{G} \implies G[h_1(\cdot),\ldots,h_m(\cdot)]\in C^{\infty}(\mathbb{M})~~~ \forall~(h_1,\dots,h_m)\in (C^{\infty}(\mathbb{M}))^m.
\end{equation}
Note that the variations of function-valued functionals are precisely the variations used in Hamilton's principle in classical field theory; see \cref{EulerLagrangeEquationsSection} for further details.

Functional derivatives and variations of function-valued functionals may be defined in a manner similar to that of functionals. This becomes clear if one considers the \textit{function-valued functional} $G[h_1(\cdot),\ldots,h_m(\cdot)](\circ)\in \mathscr{G}$ as a different \textit{functional} $G[h_1(\cdot),\ldots,h_m(\cdot)](x)\in \mathscr{F}$ for each fixed $x\in\mathbb{M}$. Note the use of the $(\circ)$ notation to indicate that the function-valued functional $G$ takes arguments from $\mathbb{M}$, and these arguments are to be distinguished from the arguments of the $h_i$ functions. We refer the reader to appendix \ref{ExamplesOfFunctionalFormalism} for examples of how the function-valued functional notation is applied. 

In what follows, we will need to consider the variations and derivatives of the constraints $g$ and $f^\nu$ as well as the variations and derivatives of the dependent field component $A^{\mu_D}$. Since $g,f^\nu$ and $A^{\mu_D}$ are dependent on the independent field components $A^{\mu}$ and their derivatives $\partial_\nu A^\mu$, with $\nu,\mu\neq \mu_D$, the variations and derivatives of $g,f^\nu$ and $A^{\mu_D}$ will induce variations on the independent field components and their derivatives. The precise form of these induced variations and derivatives will arise through chain rule type relations for function-valued functionals. The chain rule relations can be proven as a direct consequence of the definitions provided in \cref{functionalDerivativeDefinition,variationDefinition}, the proofs of which are presented in appendix \ref{ChainRuleProofs}. 

Here, we shall state the relevant chain rule relations required for our proof of the transposition rule: Consider the function-valued functional $G\in \mathscr{G}$, and suppose we compose $G$ with a set of function-valued functionals $\{G^j\in  \mathscr{G}: j=1,\ldots,m\}$; the following chain rule relations then hold: 
\begin{subequations}
    \begin{align}
        \label{functionalDerivativeChainRule}
        \frac{\partial G(x)}{\partial h_i(y)}
        &=
        \sum_{j=1}^{m} \int d^d z~
        \frac{\partial G(x)}{\partial G^j(z)}
        \frac{\partial G^j(z)}{\partial h_i(y)}\\
        \label{variationChainRule}
        \delta G(x)
        &=
        \sum_{j=1}^{m} \int d^d z~
        \frac{\partial G(x)}{\partial G^j(z)}
        \delta G^j(z)\\
        \label{derivativeChainRule}
        \partial^x_\nu G(x)
        &=
        \sum_{j=1}^{m} \int d^d z~
        \frac{\partial G(x)}{\partial G^j(z)}
        \partial^z_\nu G^j(z),
    \end{align}
\end{subequations}
where we have used the following shorthand notation:
\begin{subequations}
    \label{shorthandNotations}
    \begin{align}
        \label{GxDefinition}
        G(x)
        &\equiv 
        G[\{G^j[h_1(\cdot),\ldots,h_n(\cdot)](\circ): j=1,\ldots,m\}](x)\\
        \label{functionalDerivativeShorthand}
        \frac{\partial G(x)}{\partial h_i(y)}
        &\equiv
        \frac{\partial G[\{G^j[h_1(\cdot),\ldots,h_i(\cdot)+\alpha \delta^d(\cdot-y),\ldots,h_n(\cdot)](\circ): j=1,\ldots,m\}](x)}{\partial \alpha}\Bigg|_{\alpha=0}\\
        \label{functionalDerivativeShorthand2}
        \frac{\partial G(x)}{\partial G^j(z)}
        &\equiv
        \frac{\partial G[G^1[h_1(\cdot),\ldots,h_n(\cdot)](\circ),\ldots,G^j[h_1(\cdot),\ldots,h_n(\cdot)](\circ)+\alpha \delta^d(\circ-z),\ldots,G^m[h_1(\cdot),\ldots,h_n(\cdot)](\circ)](x)}{\partial \alpha}\Bigg|_{\alpha=0}\\
        \label{variationShortHand}
        \delta G(x)
        &\equiv
        \sum_{i=1}^{n}
        \bigg\{
        G[\{G^j[h_1(\cdot),\ldots,h_i(\cdot)+\alpha \gamma_i(\cdot),\ldots,h_n(\cdot)](\circ): j=1,\ldots,m\}](x)\\
        \notag
        &\quad~~~~-
        G[\{G^j[h_1(\cdot),\ldots,h_i(\cdot),\ldots,h_n(\cdot)](\circ): j=1,\ldots,m\}](x)
        \bigg\}
        \\
        \label{spacetimeDerivativeShorthand}
        \partial^x_\nu 
        &\equiv
        \frac{\partial}{\partial x^\nu}.
    \end{align}
\end{subequations}
In \cref{shorthandNotations}, we use the following notation
\begin{equation}
    G[\{G^j[h_1(\cdot),\ldots,h_n(\cdot)](\circ): j=1,\ldots,m\}](x)
    =
    G[G^1[h_1(\cdot),\ldots,h_n(\cdot)](\circ),\ldots,G^m[h_1(\cdot),\ldots,h_n(\cdot)](\circ)](x),
\end{equation}
while in \cref{spacetimeDerivativeShorthand}, we use $\partial_\nu^x$ to denote differentiation with respect to the spacetime coordinates $x^\nu$ which is to be distinguished from $\partial_\nu^z$ which denotes differentiation with respect to the coordinates $z^\nu$. One should note that the chain rule for the spacetime derivative in \cref{derivativeChainRule} is only valid when $G$ has the following translational property
\begin{subequations}
    \label{TranslationalPropertyOfG}
    \begin{align}
        G[h_1(\cdot+y),\ldots,h_n(\cdot+y)](x)
        &=
        G[h_1(\cdot),\ldots,h_n(\cdot)](x+y),
    \end{align}
\end{subequations}
the restriction of which is not a problem for our purposes as all the function-valued functionals which we will take spacetime derivatives of in our proof of the transposition rule satisfy \cref{TranslationalPropertyOfG}. We would like to point out that the function-valued functional $G$ is dependent on the $m\in\mathbb{N}$ function-valued functionals $G^j$, while we have allowed each of the $G^j$ function-valued functionals to depend on $n\in\mathbb{N}$ functions $h_i$, where $m$ and $n$ need not be equal. In principle, each of the $G^j$ function-valued functionals may depend on a different number of functions $n_j$; we take $n$ to be the maximum of all $n_j$ for notational simplicity.

\subsection{Proof of the Transposition Rule for Independent Components}
To prove the transposition rule for the independent field components $A^{\mu}$, with $\mu\neq \mu_D$, we consider the variation of the identity map $\mathds{1}\in\mathscr{G}$ acting on the independent field components $A^{\mu}$:
\begin{subequations}
    \label{variationOfIndependentComponent}
    \begin{align}
        \delta A^{\mu}(x)
        &=
        \delta \big(\mathds{1}[A^{\mu}(\cdot)](x)\big)\\
        \label{usingVariationDefinitionInVariationOfIndependentComponent}
        &=
        \mathds{1}[A^{\mu}(\cdot)+\alpha\gamma^\mu(\cdot)](x)
        -\mathds{1}[A^{\mu}(\cdot)](x)\\
        &=
        (A^{\mu}(x)+\alpha\gamma^\mu(x)) - A^{\mu}(x)\\
        &=
        \alpha \gamma^\mu(x),
    \end{align}
\end{subequations}
where we have made use of \cref{variationDefinition} in \cref{usingVariationDefinitionInVariationOfIndependentComponent}. Now, if one considers the derivative of the independent component $A^{\mu}$, as a function in $C^{\infty}(\mathbb{M})$, then similarly to \cref{variationOfIndependentComponent}, one can compute the variation of the derivative $\partial_\nu A^{\mu}$:
\begin{subequations}
    \begin{align}
        \label{provingTPRForIndep}
        \delta(\partial_\nu^x A^{\mu}(x))
        &=
        \delta \big(\partial_\nu^x \mathds{1}[ A^{\mu}(\cdot)](x)\big)\\
        \label{usingVariationDefinitionInVariationOfIndependentComponent2}
        &=
        \partial_\nu^x \mathds{1}[ A^{\mu}(\cdot)+\alpha\gamma^\mu(\cdot)](x)
        -\partial_\nu^x \mathds{1}[ A^{\mu}(\cdot)](x)\\
        &=
        \partial_\nu^x (A^{\mu}(x)+\alpha\gamma^\mu(x)) - \partial_\nu^x A^{\mu}(x)\\
        &=
        \alpha \partial_\nu^x \left(\gamma^\mu(x)\right)\\
        &=
        \partial_\nu^x(\delta A^{\mu}(x)),
    \end{align}
\end{subequations}
where we have again made use of \cref{variationDefinition} in \cref{usingVariationDefinitionInVariationOfIndependentComponent2}. Note that in both \cref{variationOfIndependentComponent} and \cref{provingTPRForIndep} we have made use of the definition of the variation in \cref{variationDefinition}, and
have not made use of the chain rule relation in \cref{variationChainRule}; we make this choice because the identity map $\mathds{1}$ is trivial enough that the definition of the variation in \cref{variationDefinition} may be applied directly.

The results of \cref{provingTPRForIndep} show that the transposition rule holds for the $d-1$ independent field components $A^{\mu}$, with $\mu\neq \mu_D$, and thus we may write
\begin{equation}
    \label{TranspositionRuleIndependentComponent}
    \delta(\partial_\nu A^{\mu})=\partial_\nu(\delta A^{\mu}), \quad \forall \mu\neq \mu_D.
\end{equation}

\subsection{Proof of the Transposition Rule for the Dependent Component}
\label{DependentComponentTPRProofSection}
The proof of the transposition rule for the dependent field component $A^{\mu_D}$ is more intricate than the independent case due to the functional dependence that $A^{\mu_D}$ has on the derivatives of the independent field components $\partial_\nu A^{\mu}$ with $\nu,\mu\neq\mu_D$; we will thus have to pay close attention to the chain rule relations derived in appendix \ref{ChainRuleProofs}.

We will prove the transposition rule, $\delta (\partial_\nu A^{\mu_D})=\partial_\nu (\delta A^{\mu_D})$, for the dependent field component in two separate cases: first for $\nu=\mu_D$ and second for $\nu\neq\mu_D$. In both cases we will need the variation $\delta A^{\mu_D}$, which can be calculated using the definition of the variation from \cref{variationDefinition} as follows:
\begin{subequations}
    \begin{align}
        \delta A^{\mu_D}(x)
        &=
        \sum_{(\rho,\sigma)\neq\mu_D} 
        \left(
            \frac
            {\partial A^{\mu_D}[\ldots,\partial_\rho A^\sigma(\cdot)+\alpha\gamma^\sigma_\rho(\cdot),\ldots](x)}
            {\partial \alpha}
        \right)\bigg|_{\alpha=0}\\
        \label{usingBInVariationOfDependentField}
        &=
        \sum_{(\rho,\sigma)\neq\mu_D}
        \frac
            {\partial}
            {\partial \alpha} 
        \left(
            \int_{-\infty}^{x^{\mu_D}} d\tilde{x}^{\mu_D}~
            B[\ldots,\partial_\rho A^\sigma(\cdot)+\alpha\gamma^\sigma_\rho(\cdot),\ldots](x^0,\ldots,\tilde{x}^{\mu_D},\ldots,x^{d-1})
        \right)\bigg|_{\alpha=0}\\
        \label{usingBInVariationOfDependentField2}
        &=
        \int_{-\infty}^{x^{\mu_D}} d\tilde{x}^{\mu_D}~
        \delta B(x^0,\ldots,\tilde{x}^{\mu_D},\ldots,x^{d-1}),
    \end{align}
\end{subequations}
where in \cref{usingBInVariationOfDependentField}, we have used the expression of $A^{\mu_D}$ from \cref{explicitFromOfDependentFieldComponenInTermsOfBFunctional}.

\subsection*{Case 1: \texorpdfstring{$\nu=\mu_D$}{nu equal to muD}}
If we act $\partial_{\mu_D}$ on our expression for $\delta A^{\mu_D}$ derived in \cref{usingBInVariationOfDependentField2}, we find---by a simple application of the fundamental theorem of calculus---the following
\begin{equation}
    \partial^x_{\mu_D}(\delta A^{\mu_D}(x))
    =
    \partial^x_{\mu_D}
    \left(
        \int_{-\infty}^{x^{\mu_D}} d\tilde{x}^{\mu_D}~
        \delta B(x^0,\ldots,\tilde{x}^{\mu_D},\ldots,x^{d-1})
    \right)
    =\delta B(x).
\end{equation}
Now, acting $\partial_{\mu_D}$ and then $\delta$ on $A^{\mu_D}$, we can again make use of the fundamental theorem of calculus to find
\begin{subequations}
    \begin{align}
        \delta(
        \partial^x_{\mu_D} A^{\mu_D}(x)
        )
        &=
        \delta
        \left(
            \partial^x_{\mu_D}
            \left(
                \int_{-\infty}^{x^{\mu_D}} d\tilde{x}^{\mu_D}~
                B(x^0,\ldots,\tilde{x}^{\mu_D},\ldots,x^{d-1})
            \right)
        \right)\\
        &=
        \delta B(x).
    \end{align}
\end{subequations}
Therefore, we have shown that $\delta (\partial_{\mu_D} A^{\mu_D})=\partial_{\mu_D} (\delta A^{\mu_D})$.

\subsection*{Case 2: \texorpdfstring{$\nu\neq\mu_D$}{nu not equal to muD}}
In this case, we will start by acting $\partial_\nu$ on the dependent field component $A^{\mu_D}$, in doing so we find:
\begin{subequations}
    \begin{align}
        \partial^x_{\nu} A^{\mu_D}(x)
        &=
        \partial^x_\nu
        \left(
            \int_{-\infty}^{x^{\mu_D}} d\tilde{x}^{\mu_D}~
            B(x^0,\ldots,\tilde{x}^{\mu_D},\ldots,x^{d-1})
        \right)\\
        \label{bringDerivativeThroughIntegral}
        &=
        \int_{-\infty}^{x^{\mu_D}} d\tilde{x}^{\mu_D}~
            \partial^x_\nu B(x^0,\ldots,\tilde{x}^{\mu_D},\ldots,x^{d-1}),
    \end{align}
\end{subequations}
where in \cref{bringDerivativeThroughIntegral}, the integral is over the $x^{\mu_D}$ component, so we could bring the $\partial_\nu$ derivative through the integral as $\nu\neq\mu_D$. We may now make use of the expression derived in \cref{spaceTimeDerivativeOfB} to write
\begin{equation}
    \partial^x_{\nu} A^{\mu_D}(x)
    =
    \int_{-\infty}^{x^{\mu_D}} d\tilde{x}^{\mu_D}~
    \sum_{(\rho,\sigma)\neq(\mu_D)}
    \int d^dy~
    \frac{\partial B(x^0,\ldots,\tilde{x}^{\mu_D},\ldots,x^{d-1})}{\partial (\partial^y_\rho A^\sigma(y))}
    \partial^y_\nu(\partial^y_\rho A^\sigma(y)). 
\end{equation}
Next, we act $\delta$ on $\partial_\nu A^{\mu_D}$, and make use of the product rule for variations from appendix \ref{ProductRuleForVariationsProof} to find:
\begin{subequations}
    \begin{align}
        \delta( \partial^x_{\nu} A^{\mu_D}(x))
        &=
        \delta
        \left(
            \sum_{(\rho,\sigma)\neq(\mu_D)}
            \int_{-\infty}^{x^{\mu_D}} d\tilde{x}^{\mu_D}~
            \int d^dy~
            \frac{\partial B(x^0,\ldots,\tilde{x}^{\mu_D},\ldots,x^{d-1})}{\partial (\partial^y_\rho A^\sigma(y))}
            \partial^y_\nu(\partial^y_\rho A^\sigma(y)) 
        \right)\\
        &=
            \sum_{(\rho,\sigma)\neq(\mu_D)}
            \int_{-\infty}^{x^{\mu_D}} d\tilde{x}^{\mu_D}~
            \int d^dy~
        \delta
        \left(
            \frac{\partial B(x^0,\ldots,\tilde{x}^{\mu_D},\ldots,x^{d-1})}{\partial (\partial^y_\rho A^\sigma(y))}
        \right)
            \partial^y_\nu(\partial^y_\rho A^\sigma(y))
        \\
        \notag
        &+
            \sum_{(\rho,\sigma)\neq(\mu_D)}
            \int_{-\infty}^{x^{\mu_D}} d\tilde{x}^{\mu_D}~
            \int d^dy~
            \frac{\partial B(x^0,\ldots,\tilde{x}^{\mu_D},\ldots,x^{d-1})}{\partial (\partial^y_\rho A^\sigma(y))}
        \delta
        \bigg(
            \partial^y_\nu(\partial^y_\rho A^\sigma(y)) 
        \bigg)\\
        \label{usingvariationOfFunctionalDerivativeOfBFinal}
        &=
            \sum_{(\rho,\sigma)\neq(\mu_D)}
            \int_{-\infty}^{x^{\mu_D}} d\tilde{x}^{\mu_D}~
            \int d^dy~ 
        \bigg(
            \overbrace{   
            \sum_{(\rho^\prime,\sigma^\prime)\neq(\mu_D)}
            \int d^dz~
            \frac
            {\partial^2 B(x^0,\ldots,\tilde{x}^{\mu_D},\ldots,x^{d-1})}
            {\partial (\partial^z_{\rho^\prime}A^{\sigma^\prime}(z)) \partial (\partial^y_\rho A^\sigma(y))}
            \delta(\partial^z_{\rho^\prime}A^{\sigma^\prime}(z))
            }^{\cref{variationOfFunctionalDerivativeOfBFinal}}
        \bigg)
        \partial^y_\nu(\partial^y_\rho A^\sigma(y))
        \\
        \notag
        &+
            \sum_{(\rho,\sigma)\neq(\mu_D)}
            \int_{-\infty}^{x^{\mu_D}} d\tilde{x}^{\mu_D}~
            \int d^dy~
            \frac{\partial B(x^0,\ldots,\tilde{x}^{\mu_D},\ldots,x^{d-1})}{\partial (\partial^y_\rho A^\sigma(y))}
        \delta
        \bigg(
            \partial^y_\nu(\partial^y_\rho A^\sigma(y)) 
        \bigg),
    \end{align}
\end{subequations}
where in \cref{usingvariationOfFunctionalDerivativeOfBFinal}, we have used the expression derived in \cref{variationOfFunctionalDerivativeOfBFinal}. Next, we will rearrange the first expression in \cref{usingvariationOfFunctionalDerivativeOfBFinal}, to make use of the expression derived in \cref{secondOrderDerivativeOfBFinal2}.
\begin{subequations}
    \begin{align}
        \delta( \partial^x_{\nu} A^{\mu_D}(x))
        &=
            \int_{-\infty}^{x^{\mu_D}} d\tilde{x}^{\mu_D}~
            \sum_{(\rho^\prime,\sigma^\prime)\neq(\mu_D)}
            \int d^dz~
        \left(
            \sum_{(\rho,\sigma)\neq(\mu_D)}
            \int d^dy~
            \frac
            {\partial^2 B(x^0,\ldots,\tilde{x}^{\mu_D},\ldots,x^{d-1})}
            {\partial (\partial^z_{\rho^\prime}A^{\sigma^\prime}(z)) \partial (\partial^y_\rho A^\sigma(y))}
            \partial^y_\nu(\partial^y_\rho A^\sigma(y))
        \right)
        \delta
        \bigg(
            \partial^z_{\rho^\prime}A^{\sigma^\prime}(z)
        \bigg)
        \\
        \notag
        &+
            \sum_{(\rho,\sigma)\neq(\mu_D)}
            \int_{-\infty}^{x^{\mu_D}} d\tilde{x}^{\mu_D}~
            \int d^dy~
            \frac{\partial B(x^0,\ldots,\tilde{x}^{\mu_D},\ldots,x^{d-1})}{\partial (\partial^y_\rho A^\sigma(y))}
        \delta
        \bigg(
            \partial^y_\nu(\partial^y_\rho A^\sigma(y)) 
        \bigg)\\
        &\overset{\cref{secondOrderDerivativeOfBFinal2}}{=}
            \sum_{(\rho^\prime,\sigma^\prime)\neq(\mu_D)}
            \int_{-\infty}^{x^{\mu_D}} d\tilde{x}^{\mu_D}~
            \int d^dz~
            \partial_\nu^x 
            \left(
                \frac{\partial B(x^0,\ldots,\tilde{x}^{\mu_D},\ldots,x^{d-1})}{\partial (\partial^z_{\rho^\prime} A^{\sigma^\prime}(z))}
            \right)
        \delta
        \bigg(
            \partial^z_{\rho^\prime}A^{\sigma^\prime}(z)
        \bigg)
        \\
        \notag
        &\qquad +
            \sum_{(\rho^\prime,\sigma^\prime)\neq(\mu_D)}
            \int_{-\infty}^{x^{\mu_D}} d\tilde{x}^{\mu_D}~
            \int d^dz~
            \partial^z_\nu
            \left(
                \frac{\partial B(x^0,\ldots,\tilde{x}^{\mu_D},\ldots,x^{d-1})}{\partial (\partial^z_{\rho^\prime} A^{\sigma^\prime}(z))}
            \right)
        \delta
        \bigg(
            \partial^z_{\rho^\prime}A^{\sigma^\prime}(z)
        \bigg)
        \\
        \notag
        &\qquad +
            \sum_{(\rho,\sigma)\neq(\mu_D)}
            \int_{-\infty}^{x^{\mu_D}} d\tilde{x}^{\mu_D}~
            \int d^dy~
            \frac{\partial B(x^0,\ldots,\tilde{x}^{\mu_D},\ldots,x^{d-1})}{\partial (\partial^y_\rho A^\sigma(y))}
        \delta
        \bigg(
            \partial^y_\nu(\partial^y_\rho A^\sigma(y)) 
        \bigg)\\
        \label{useIntegrationByPartsToMoveDerivativeOffOfFunctionalDerivativeOfB}
        &\overset{\text{IBP}}{=}
            \sum_{(\rho^\prime,\sigma^\prime)\neq(\mu_D)}
            \int_{-\infty}^{x^{\mu_D}} d\tilde{x}^{\mu_D}~
            \int d^dz~
            \partial_\nu^x 
            \left(
                \frac{\partial B(x^0,\ldots,\tilde{x}^{\mu_D},\ldots,x^{d-1})}{\partial (\partial^z_{\rho^\prime} A^{\sigma^\prime}(z))}
            \right)    
        \delta
        \bigg(
            \partial^z_{\rho^\prime}A^{\sigma^\prime}(z)
        \bigg)
        \\
        \notag
        &\qquad -
            \sum_{(\rho^\prime,\sigma^\prime)\neq(\mu_D)}
            \int_{-\infty}^{x^{\mu_D}} d\tilde{x}^{\mu_D}~
            \int d^dz~
            \frac{\partial B(x^0,\ldots,\tilde{x}^{\mu_D},\ldots,x^{d-1})}{\partial (\partial^z_{\rho^\prime} A^{\sigma^\prime}(z))}
            \partial^z_\nu
            \left(
                \delta
                \bigg(
                    \partial^z_{\rho^\prime}A^{\sigma^\prime}(z)
                \bigg)
            \right)
        \\
        \notag
        &\qquad +
            \sum_{(\rho,\sigma)\neq(\mu_D)}
            \int_{-\infty}^{x^{\mu_D}} d\tilde{x}^{\mu_D}~
            \int d^dy~
            \frac{\partial B(x^0,\ldots,\tilde{x}^{\mu_D},\ldots,x^{d-1})}{\partial (\partial^y_\rho A^\sigma(y))}
        \delta
        \bigg(
            \partial^y_\nu(\partial^y_\rho A^\sigma(y)) 
        \bigg),
    \end{align}
\end{subequations}
where in \cref{useIntegrationByPartsToMoveDerivativeOffOfFunctionalDerivativeOfB}, we have used integration by parts to move the $\partial_\nu^z$ derivative off of the functional derivative of $B$. In \cref{TranspositionRuleIndependentComponent}, we showed that the transposition rule for the independent components, $\delta(\partial_\nu A^\mu)=\partial_\nu(\delta A^\mu)~\forall \mu\neq\mu_D$, holds. This proof followed trivially since the field components $A^\mu~\forall \mu\neq\mu_D$ were allowed to vary independently. Recall, that we have the expression, 
$\partial_{\mu_D}A^{\mu_D}=-\sum_{\nu\neq\mu_D} \partial_\nu f^\nu [A^\mu:\mu\neq\mu_D](x)$. Hence, we see that the only dependent derivative is $\partial_{\mu_D}A^{\mu_D}$, and all the other derivatives $\partial_\nu A^\mu~\forall \nu,\mu\neq\mu_D$ are independent. Thus, we may conclude that the derivatives $\partial_\nu A^\mu~\forall \nu,\mu\neq\mu_D$ will vary independently, and therefore the variation $\delta$ will commute with the derivatives of the derivatives of the independent field components, \textit{i.e.} $\delta(\partial_\nu(\partial_\rho A^\mu))=\partial_\nu(\delta(\partial_\rho A^\mu))$ for all $\mu\neq\mu_D$. Returning to our calculation, one will notice---after commuting the variation $\delta$ past the $\partial_\nu$ spacetime-derivative in the second term of \cref{useIntegrationByPartsToMoveDerivativeOffOfFunctionalDerivativeOfB}---that the second and third terms cancel in \cref{useIntegrationByPartsToMoveDerivativeOffOfFunctionalDerivativeOfB}. Thus, we have
\begin{subequations}
    \begin{align}
        \label{pullOutSpaceTimeDerivativeOfIntegral}
        \delta( \partial^x_{\nu} A^{\mu_D}(x))
        &=
        \partial_\nu^x 
        \left(
            \int_{-\infty}^{x^{\mu_D}} d\tilde{x}^{\mu_D}~
            \sum_{(\rho^\prime,\sigma^\prime)\neq(\mu_D)}
            \int d^dz~
            \frac{\partial B(x^0,\ldots,\tilde{x}^{\mu_D},\ldots,x^{d-1})}{\partial (\partial^z_{\rho^\prime} A^{\sigma^\prime}(z))}
            \delta(\partial^z_{\rho^\prime}A^{\sigma^\prime}(z))
        \right)\\
        &=
        \label{useVariationOfB}
        \partial_\nu^x 
        \left(
            \int_{-\infty}^{x^{\mu_D}} d\tilde{x}^{\mu_D}~
            \delta B(x^0,\ldots,\tilde{x}^{\mu_D},\ldots,x^{d-1})
        \right)
    \end{align}
\end{subequations}
where in \cref{pullOutSpaceTimeDerivativeOfIntegral}, we pulled out the $\partial_\nu$ spacetime derivative from the first term in \cref{useIntegrationByPartsToMoveDerivativeOffOfFunctionalDerivativeOfB}. In \cref{useVariationOfB}, we used the chain rule for variations from \cref{variationChainRule}. Our final movement in the proof of the transposition rule, will be to recognize the term in the parenthesis of \cref{useVariationOfB} as the variation of the dependent field component, which was derived in \cref{usingBInVariationOfDependentField2}. We may thus write
\begin{equation}
    \delta( \partial^x_{\nu} A^{\mu_D}(x))=\partial_\nu^x(\delta A^{\mu_D}).
\end{equation}

Hence, we may finally conclude our proof of the transposition rule for all field components $A^{\mu}$ and therefore write:
\begin{equation}
    \label{finalTranspositionRule}
    \boxed{
        \phantom{\strut}
        \delta (\partial_\nu  A^\mu) =\partial_\nu(\delta  A^\mu) ~~~ \forall \nu ~\&~ \forall \mu
        \phantom{\strut}
        }
\end{equation}

\section{Classical Field Theory with Constraints}
\label{ClassicalFieldTheoryWithConstraintsSection}
In this section we shall apply the formalism developed in \cref{ProofOfTranspositionRuleSection} along with the transposition rule derived therein to classical field theories with integrable non-holonomic constraints. We shall first derive the Euler-Lagrange equations for classical field theories with integrable non-holonomic constraints in \cref{EulerLagrangeEquationsSection}, before then applying the formalism to the specific example of classical electrodynamics in \cref{ClassicalElectrodynamicsSection}.

\subsection{Hamilton's Principle and the Euler-Lagrange Equations} 
\label{EulerLagrangeEquationsSection}
Consider a classical field theory in $d$ spacetime dimensions described by the Lagrangian density $\mathscr{L}$ with $d$ field components $A^\mu$, and suppose that the theory is placed under an integrable non-holonomic constraint
\begin{equation}
    \label{ConstraintDef}
    g[\partial_\nu A^\mu](x)=0.
\end{equation}
We will require the variations to respect the constraint in \cref{ConstraintDef}, \textit{i.e.}
\begin{equation}
    \label{variationsRespectConstraint}
    g[\partial_\nu A^\mu+\delta(\partial_\nu A^\mu)](x)=0.
\end{equation}
\Cref{variationsRespectConstraint} then implies the following
\begin{subequations}
    \begin{align}
        0
        &=
        g[\partial_\nu A^\mu+\delta(\partial_\nu A^\mu)](x)\\
        &=
        \overbrace{g[\partial_\nu A^\mu](x)}^{0}
        +
        \sum_{\mu,\nu=0}^{d-1} \int d^d y~
        \frac{\partial g(x)}{\partial (\partial_\nu^y A^{\mu}(y))}
        ~\delta(\partial_\nu^y A^{\mu}(y))\\
        \label{variationOfGvanishes}
        \implies
        0
        &=    
        \sum_{\mu,\nu=0}^{d-1} \int d^d y~
        \frac{\partial g(x)}{\partial (\partial_\nu^y A^{\mu}(y))}
        ~\delta(\partial_\nu^y A^{\mu}(y))
    \end{align}
\end{subequations}
where we have expanded to first order in the variation $\delta$. Multiplying both sides of \cref{variationOfGvanishes} by an arbitrary function $\Lambda(x)$---that is independent of the $A^\mu$ field components and their derivatives---yields
\begin{equation}
    0=\sum_{\mu,\nu=0}^{d-1} \int d^d y~
    \frac{\partial }{\partial (\partial_\nu^y A^{\mu}(y))}
    \big[\Lambda(x)  g(x)\big]
    ~\delta(\partial_\nu^y A^{\mu}(y)).
\end{equation}
The $\Lambda$ function will perform the role of a Lagrange multiplier \cite{goldstein:mechanics,Henneaux:1992ig,taylor2005classical,Bloch2015Nonholonomic}. We may now vary the action $S$---which is dependent on the field components $A^\mu$ and their derivatives $\partial_\nu A^\mu$ which are varied independently from each other---and through Hamilton's principle\cite{Hamilton1834}, set the variation to zero:
\begin{subequations}
    \begin{align}
        0&=\delta S[A^\mu,\partial_\nu A^\mu]\\
        &=
        \int d^d x~
        \delta \mathscr{L}[A^\mu,\partial_\nu A^\mu](x)\\
        &=
        \label{usingVariationChainRuleOnLagrangian}
        \int d^d x~ d^dy
        \left(
            \sum_{\mu=0}^{d-1}
            \frac{\partial \mathscr{L}(x)}{\partial A^{\mu}(y)}
            ~\delta A^{\mu}(y)
            +
            \sum_{\mu,\nu=0}^{d-1}
            \frac{\partial \mathscr{L}(x)}{\partial (\partial_\nu^y A^{\mu}(y))}
            ~\delta(\partial_\nu^y A^{\mu}(y))
        \right)\\
        \label{addingConstraintVariationToActionVariation}
        &=
        \int d^d x~ d^dy
        \left(
            \sum_{\mu=0}^{d-1}
            \frac{\partial }{\partial A^{\mu}(y)}
            \big[
                \mathscr{L}(x)
                +
                \Lambda(x)  g(x)
            \big]
            ~\delta A^{\mu}(y)
            +
            \sum_{\mu,\nu=0}^{d-1}
            \frac{\partial }{\partial (\partial_\nu^y A^{\mu}(y))}
            \big[
                \mathscr{L}(x)
                +
                \Lambda(x)  g(x)
            \big]
            ~\delta(\partial_\nu^y A^{\mu}(y)),
        \right)
    \end{align}
\end{subequations}
in \cref{usingVariationChainRuleOnLagrangian} we have applied the chain rule for variations in \cref{variationChainRule} to the Lagrangian density $\mathscr{L}$. In \cref{addingConstraintVariationToActionVariation}, we have added two zeros in the form of the variation of the constraint multiplied by the Lagrange multiplier $\Lambda(x)$ to the variation of the action: The first zero arises as both $g$ and $\Lambda$ are independent of the $A^\mu$ field components; thus, $\partial (\Lambda g)/\partial A^\mu=0$. The second zero comes from \cref{variationOfGvanishes}.

Next we shall make use of the transposition rule in \cref{finalTranspositionRule} and perform an integration by parts on the second term in \cref{addingConstraintVariationToActionVariation}:
\begin{subequations}
    \begin{align}
        \delta S
        &=
        \int d^d x~ d^dy
        \left(
            \sum_{\mu=0}^{d-1}
            \frac{\partial }{\partial A^{\mu}(y)}
            \big[
                \mathscr{L}(x)
                +
                \Lambda(x)  g(x)
            \big]
            ~\delta A^{\mu}(y)
            +
            \sum_{\mu,\nu=0}^{d-1}
            \frac{\partial }{\partial (\partial_\nu^y A^{\mu}(y))}
            \big[
                \mathscr{L}(x)
                +
                \Lambda(x)  g(x)
            \big]
            ~\partial_\nu^y(\delta A^{\mu}(y))
        \right)\\
        &=
        \int d^d x~ d^dy
        \left(
            \sum_{\mu=0}^{d-1}
            \frac{\partial }{\partial A^{\mu}(y)}
            \big[
                \mathscr{L}(x)
                +
                \Lambda(x)  g(x)
            \big]
            -
            \sum_{\mu,\nu=0}^{d-1}
            \partial_\nu^y
            \bigg(
                \frac{\partial }{\partial (\partial_\nu^y A^{\mu}(y))}
                \big[
                    \mathscr{L}(x)
                    +
                    \Lambda(x)  g(x)
                \big]
            \bigg)
        \right)
        ~\delta A^{\mu}(y)
    \end{align}
\end{subequations}
Now, the integrable non-holonomic constraint $g$ induces a functional dependence between the field components and their variations, as discussed in \cref{ProofOfTranspositionRuleSection}. Thus, the variations $\delta A^{\mu}$ are not all independent; rather the $\delta {A^{\mu_D}}$ variation will depend functionally on the independent variations $\delta A^{\mu}$, with $\mu\neq \mu_D$. Therefore, we may not set the integrand of the variation of the action to zero for each field component independently---as is ordinarily done when deriving the Euler-Lagrange equations for unconstrained dynamical systems. Instead, the argument goes as follows: Choose $\Lambda$, the Lagrange multiplier, such that
\begin{equation}
    \label{dependentEulerLagrange}
    \int d^dx
        \left(
            \frac{\partial }{\partial A^{\mu_D}(y)}
            \big[
                \mathscr{L}(x)
                +
                \Lambda(x)  g(x)
            \big]
            -
            \sum_{\nu=0}^{d-1}
            \partial_\nu^y
            \bigg(
                \frac{\partial }{\partial (\partial_\nu^y A^{\mu_D}(y))}
                \big[
                    \mathscr{L}(x)
                    +
                    \Lambda(x)  g(x)
                \big]
            \bigg)
        \right)=0.
\end{equation}
The remaining $d-1$ field variations $\delta A^{\mu}$, with $\mu\neq\mu_D$, are all independent; thus, by the fundamental theorem of the calculus of variations\cite{GelfandFomin2000}, Hamilton's principle (\textit{i.e.} $\delta S=0$) implies that
\begin{equation}
    \label{independentEulerLagrange}
    \int d^dx
        \left(
            \frac{\partial }{\partial A^{\mu}(y)}
            \big[
                \mathscr{L}(x)
                +
                \Lambda(x)  g(x)
            \big]
            -
            \sum_{\nu=0}^{d-1}
            \partial_\nu^y
            \bigg(
                \frac{\partial }{\partial (\partial_\nu^y A^{\mu}(y))}
                \big[
                    \mathscr{L}(x)
                    +
                    \Lambda(x)  g(x)
                \big]
            \bigg)
        \right)=0, \quad \mu\neq \mu_D.
\end{equation}
Supplemented with initial conditions, \cref{dependentEulerLagrange} yields one equation to solve for $\Lambda$, while \cref{independentEulerLagrange} provides an additional $d - 1$ equations to determine the independent $A^{\mu}$ components. Once these independent components have been determined, the dependent component $A^{\mu_D}$ can be obtained through its functional dependence on the independent components, thereby solving the dynamical system.

Upon closer inspection, one may notice that this procedure of including constraints into the classical field theory is mathematically equivalent to adjoining the constraints weighted by the Lagrange multipliers to the original Lagrangian; \textit{i.e.} 
\begin{equation}
     \label{LagrangianWithConstraint}
     \mathscr{L}\mapsto \mathscr{L}_{\text{adjoined}}\equiv\mathscr{L}+\overbrace{\mathscr{L}_{\text{constraint}}}^{\Lambda g},
\end{equation}
and then solving the usual Euler-Lagrange equations supplemented with the constraint equation in \cref{ConstraintDef}. The result in \cref{LagrangianWithConstraint}, derived in the context of classical field theory, is directly analogous to the standard result obtained when applying the method of Lagrange multipliers in the mechanics of point particles, where one similarly adjoins the Lagrangian by constraint terms weighted by multipliers\cite{taylor2005classical,goldstein:mechanics}. 

The authors would like to emphasize that the result in point particle mechanics analogous to \cref{LagrangianWithConstraint} is only valid for integrable non-holonomic constraints; if the constraint were non-integrable in point particle mechanics, then the resulting equations of motion which arise from the adjoined Lagrangian would lead to the incorrect (\textit{i.e.} in disagreement with experiment) vakonomic equations\cite{Cortes:2002abc,Bloch2015Nonholonomic,Gaset:2022ibk,Talamucci:2025abc}. We conjecture that the same is true in the context of classical field theory, and that the adjoined Lagrangian in \cref{LagrangianWithConstraint} will only yield the correct equations of motion for integrable non-holonomic constraints.

\subsection{Classical Electrodynamics with Integrable Non-Holonomic Constraints}
\label{ClassicalElectrodynamicsSection}
In this section, we demonstrate the applicability of our results to classical electrodynamics by gauge fixing the electromagnetic field using the Coulomb and Lorenz gauges. We show that both gauges are integrable non-holonomic constraints as defined in \cref{integrabilityOfConstraint}, and therefore the transposition rule in \cref{finalTranspositionRule} holds. We shall also provide an example of a non-integrable non-holonomic gauge known as the 't Hooft-Veltman (tHV) gauge. 

Classical electrodynamics may be formulated in terms of $U(1)$ gauge fields and the following Lagrangian density\cite{Jackson:1998nia,Zangwill:1507229}
\begin{equation}
\label{U1Lagrangian}
 \mathscr{L}_{U(1)}\equiv -\frac{1}{4}\sum_{\mu,\nu=0}^{3} F^{\mu \nu} F_{\mu \nu},
\end{equation}
where $F^{\mu \nu}\equiv\partial^\mu A^\nu-\partial^\nu A^\mu$ is the field strength tensor, with the metric given by $\eta_{\mu\nu}=\text{diag}(+---)_{\mu\nu}$. The $\mathscr{L}_{U(1)}$ Lagrangian density is gauge invariant and can be subjected to gauge fixing \cite{Jackson:1998nia,Jackson:2001ia,Zangwill:1507229}. The Coulomb and Lorenz gauges impose the following constraints on the $A^\mu$ gauge field:
\begin{subequations}
    \label{gaugesEq}
    \begin{align}
        g_{Lorenz}[\partial_\nu A^\mu(\cdot)](x) & \equiv \sum_{\mu=0}^3\partial_\mu A^\mu(x)=0\\
        g_{Coulomb}[\partial_\nu A^\mu(\cdot)](x) & \equiv \sum_{i=1}^{3}\partial_i A^i(x) = 0.
    \end{align}
\end{subequations}
While the tHV gauge condition is given by
\begin{equation}
    g_{tHV}[A^\mu(\cdot),\partial_\nu A^\mu(\cdot)](x) 
    \equiv 
    \sum_{\mu=0}^3
    \bigg(
    \partial_\mu A^\mu(x)+ \omega A^\mu(x)A_\mu(x)
    \bigg)
    =0,
\end{equation}
where $\omega\neq0$ is a real parameter
\cite{Koplik:1978je,Mann:1984kam,Mckeon:1985vcr,Parthasarathy:1988zf,Gracey:2005vu,Gracey:2007rz,deGracia:2019qwj,Tran:2022fdb,Phan:2023xgi}.

To match the notation used in \cref{integrabilityOfConstraint}, one may express the Lorenz gauge choice as
\begin{align}
    g_{Lorenz}[A^\mu(\cdot),\partial_\nu A^\mu(\cdot)](x)= \sum_{\nu=0}^{3}\partial_\nu f^\nu_{Lorenz}[A^\mu(\cdot)](x)=0,
\end{align}
where the function $f^\nu_{Lorenz}[A^\mu(\cdot)](x)$ is
\begin{equation}
    f^\nu_{Lorenz}[A^\mu(\cdot)](x)\equiv A^\nu(x).
\end{equation}
In the Lorenz gauge, any of the spatial components of the gauge field $A^i$ with $i=1,2,3$ may be chosen as the dependent field component $A^{\mu_D}$, with the remaining components being independent. Thus, the Lorenz gauge is an integrable, non-holonomic constraint as defined in \cref{integrabilityOfConstraint}.  The Coulomb gauge works similarly, except now the function defining the integrability condition takes the form,
\begin{equation}
    f^\nu_{Coulomb}[A^\mu(\cdot)](x)\equiv
    \begin{cases}
         0 & \text{if } \nu=0,\\
        A^\nu(x) & \text{if } \nu=1,2,3.
    \end{cases}
\end{equation}
Notice that there is no way to express the tHV gauge in the form given in \cref{integrabilityOfConstraint}. Therefore, the tHV gauge is a non-integrable, non-holonomic constraint.

Since the Coulomb and Lorenz gauges are non-holonomic constraints that satisfy the condition of integrability, one may freely adjoin the constraints in \cref{gaugesEq} to \cref{U1Lagrangian} as is commonly done in the literature \cite{Weinberg_QFT1,Weinberg_QFT2,Henneaux:1992ig,Peskin:IntroQFT,Srednicki:2007qs,Ryder_QFT,Coleman:2018mew,GaugeTheoriesOfTheStrongAndElectroweakInteraction,Zee:2003mt,Dirac1964}. Flannery\cite{Flannery:2011} showed that in the context of point particle mechanics, non-integrable non-holonomic constraints lead to a modification of the transposition rule. We expect to find an analogous result in field theory, which would have implications for gauges such as the tHV gauge. We leave this analysis for future work.

\section{Conclusions}
\label{ConclusionSection}

In this manuscript, we identified a subtlety previously unrecognized in the derivation of the equations of motion for a classical field subject to general non-holonomic constraints: the transposition rule, $\delta (\partial_\nu  A^\mu) = \partial_\nu(\delta  A^\mu)$, for such systems is expected to no longer hold in general.  
Gauge fixing in field theory imposes constraints on the field; therefore, this expected failure of the transposition rule calls into question classical  field theory results that use a gauge that depends on the derivatives of the field.

We defined integrability in the context of non-holonomic constraints in field theory and proved that the transposition rule holds for both holonomic and integrable, non-holonomic constraints. As a result, we have shown that the equations of motion for field theories with such constraints can be derived by varying the action in the usual way. 
We applied our formalism to the $U(1)$ gauge field in the Coulomb and Lorenz gauges, which are non-holonomic yet integrable as defined here.

Our results place the gauge fixing of gauge theories with non-holonomic but integrable gauges---such as the Coulomb or Lorenz gauges---on a more rigorous footing, avoiding \textit{ad hoc} methodologies, and further provide insight into the subtleties involved in gauge fixing classical and quantum gauge theories. The formalism developed in this manuscript was presented in the context of gauge fixing a classical field theory composed of a vector field $A^\mu$; however, the formalism is general and may be applied to more general classical field theories (for instance, Yang-Mills theory) with integrable non-holonomic constraints.

The future work we envisage includes performing an analysis on non-integrable, non-holonomic constraints in field theory, and exploring how/if the transposition rule must be modified in such scenarios. Often one fixes a gauge classically by adjoining the original Lagrangian with the gauge constraint multiplied by a Lagrange multiplier.  In point particle mechanics, such a procedure leads to the so-called vakonomic equations of motion; experiment has shown decisively that, when the results of vakonomic mechanics and the Lagrange-d'Alembert equations of motion differ, the Lagrange-d'Alembert equations give the correct classical path \cite{Lewis:1995vpc}.  Initial investigations of the classical $A^\mu$ solution in the 't Hooft-Veltman gauge have shown that the physical $\vec E$ and $\vec B$ fields are nevertheless reproduced by a vakonomic treatment of the gauge constraint \cite{Bert:2026fut}.  It would be very interesting to understand why it appears that following a seemingly incorrect procedure for fixing the gauge nevertheless yields the correct physical $\vec E$ and $\vec B$ fields.  Similarly, in point particle mechanics, one must include the additional terms from the modified transposition rule for non-integrable, non-holonomic constraints \cite{Flannery:2011} in order to derive the correct equations of motion from the Dirac-Bergmann algorithm \cite{Horowitz:2024eea}; thus one would expect that there are terms missing in a BRST-type gauge fixing treatment of non-integrable, non-holonomic gauges, such as the 't Hooft-Veltman gauge, which may or may not have physical consequences.

\begin{acknowledgments}
The authors would like to thank the South African National Research Foundation (NRF), the National Institute for Theoretical and Computational Sciences (NITheCS), and the SA-CERN collaboration for their generous financial support during this work. BB would like to thank Alexander Rothkopf and Eder Kikianty for their insightful comments and guidance. This research was conducted in part by WAH while visiting the Okinawa Institute of Science and Technology (OIST) through the Theoretical Sciences Visiting Program (TSVP).
\end{acknowledgments}

\appendix
\label{Appendix}
\section{Proofs of Chain Rule and Product Rule Relations for Functional Valued Functionals}
\label{ChainRuleProofs}

\subsection{Proof of the Chain Rule for Functional Derivatives}
Here, we shall prove the chain rule for functional derivatives in \cref{functionalDerivativeChainRule}. Consider the function-valued functional $G\in \mathscr{G}$, and suppose we compose $G$ with a set of function-valued functionals $\{G^j\in  \mathscr{G}: j=1,\ldots,m\}$. Then, using the definition in \cref{functionalDerivativeShorthand}, we have:
\begin{subequations}
    \begin{align}
        \frac{\partial G(x)}{\partial h_i(y)}
        &\equiv
        \frac{\partial G[\{G^j[h_1(\cdot),\ldots,h_i(\cdot)+\alpha \delta^d(\cdot-y),\ldots,h_n(\cdot)](\circ): j=1,\ldots,m\}](x)}{\partial \alpha}\Bigg|_{\alpha=0}\\
        \label{modyfyingNotation}
        &=
        \frac{\partial G[\{G^j(\alpha|i)(\circ): j=1,\ldots,m\}](x)}{\partial \alpha}\Bigg|_{\alpha=0}\\
        &=
        \label{treatingGAsFunctionOfAlpha}
        \frac
        {\partial}{\partial \alpha}
        \left(
            G
            \left[
            \left\{
                G^j(0|i)(\circ)
                +
                \alpha
                \left(
                \frac{\partial G^j(\alpha|i)(\circ)}{\partial \alpha}\Bigg|_{\alpha=0}
                \right)
                : j=1,\ldots,m
            \right\}
            \right](x)
        \right)
        \Bigg|_{\alpha=0}\\
        \label{usingOrdinaryChainRule}
        &=
        \sum_{j=1}^{m}
        \frac
        {\partial}{\partial \alpha}
        \left(
            G
            \left[
                G^1(0|i),\ldots,
                G^j(0|i)(\circ)
                +\alpha
                \left(
                \frac{\partial G^j(\alpha|i)(\circ)}{\partial \alpha}\Bigg|_{\alpha=0}
                \right),
                \ldots,
                G^m(0|i)(\circ)
            \right](x)
        \right)
        \Bigg|_{\alpha=0}
    \end{align}
\end{subequations}
where in \cref{modyfyingNotation} we have introduced the notation $G^j(\alpha|i)(\circ)\equiv G^j[h_1(\cdot),\ldots,h_i(\cdot)+\alpha \delta^d(\cdot-y),\ldots,h_n(\cdot)](\circ)$, while in \cref{treatingGAsFunctionOfAlpha} we have treated each $G^j(\alpha|i)$ as a function of the variable $\alpha$, and expanded around $\alpha=0$ to first order. Passing from \cref{treatingGAsFunctionOfAlpha} to \cref{usingOrdinaryChainRule}, we changed the $\alpha$ derivative acting on all the $G^j(\alpha|i)$ functionals simultaneously, to the $\alpha$ derivative acting on each $G^j(\alpha|i)$ functional individually. This is simply an application of the ordinary chain rule for differentiation with respect to the real variable $\alpha$. Now, recall the definition of the functional derivative in \cref{functionalDerivativeDefinition}, which we state here for an arbitrary function-valued functional $\tilde{G}$ for the sake of the reader's convenience:
\begin{equation}
    \label{functionalDerivativeDefinition2}
    \int d^d z \frac{\partial \tilde{G}(x)}{\partial h_j(z)} \gamma_j(z) \equiv \frac{\partial \tilde{G}[h_1(\cdot),\ldots,h_j(\cdot)+\alpha \gamma_j(\cdot),\ldots,h_n(\cdot)](x)}{\partial \alpha}\Bigg|_{\alpha=0},
\end{equation}
Comparing \cref{usingOrdinaryChainRule} with \cref{functionalDerivativeDefinition2}, we see that we may write:
    \begin{align}
        \label{functionalDerivativeChainRuleProofExpression}
        \frac{\partial G(x)}{\partial h_i(y)}
        &=
        \sum_{j=1}^{m}
        \int d^d z~
        \frac{\partial G(x)}{\partial G^j(0|j)(z)}
        \frac{\partial G^j(\alpha|i)(z)}{\partial \alpha}\Bigg|_{\alpha=0}
    \end{align}
Disentangling the notation, we see that $G^j(0|j)(z)=G^j[h_1(\cdot),\ldots,h_i(\cdot),\ldots,h_n(\cdot)](z)$, furthermore we have  
\begin{subequations}
    \begin{align}
        \frac{\partial G^j(\alpha|i)(z)}{\partial \alpha}\Bigg|_{\alpha=0}
        &=
        \frac{\partial G^j[h_1(\cdot),\ldots,h_i(\cdot)+\alpha \delta^d(\cdot-y),\ldots,h_n(\cdot)](z)}{\partial \alpha}\Bigg|_{\alpha=0}\\
        \label{usingDeltaFunctionDefOfFunctionalDerivative}
        &=
        \frac{\partial G^j(z)}{\partial h_i(y)},
    \end{align}
\end{subequations}
in \cref{usingDeltaFunctionDefOfFunctionalDerivative} we have used \cref{gammaAsDeltaFunction} to express the derivative with respect to $\alpha$ in terms of a functional derivative. Finally, \cref{functionalDerivativeChainRuleProofExpression} then becomes;
\begin{equation}
    \frac{\partial G(x)}{\partial h_i(y)}
    =
    \sum_{j=1}^{m}
    \int d^d z~
    \frac{\partial G(x)}{\partial G^j(z)}
    ~
    \frac{\partial G^j(z)}{\partial h_i(y)},
\end{equation}
as stated in \cref{functionalDerivativeChainRule}.

\subsection{Proof of the Chain Rule for Variations}
\label{VariationChainRuleProof}
Here we shall prove the chain rule for variations in \cref{variationChainRule}. Consider the function-valued functional $G\in \mathscr{G}$, and suppose we compose $G$ with a set of function-valued functionals $\{G^j\in  \mathscr{G}: j=1,\ldots,m\}$. Then, using the definition in \cref{variationShortHand}, we have:
\begin{subequations}
    \begin{align}
        \delta G(x)
        &\equiv
        \sum_{i=1}^{n}
        \bigg\{
        G[\{G^j[h_1(\cdot),\ldots,h_i(\cdot)+\alpha \gamma_i(\cdot),\ldots,h_n(\cdot)](\circ): j=1,\ldots,m\}](x)\\
        \notag
        &\quad~~~~-
        G[\{G^j[h_1(\cdot),\ldots,h_i(\cdot),\ldots,h_n(\cdot)](\circ): j=1,\ldots,m\}](x)
        \bigg\}\\
        \label{expandingDeltaGInAlpha}
        &=
        \sum_{i=1}^{n}
        \left(
        \frac{\partial G[\{G^j[h_1(\cdot),\ldots,h_i(\cdot)+\alpha \gamma_i(\cdot),\ldots,h_n(\cdot)](\circ): j=1,\ldots,m\}](x)}{\partial \alpha}\Bigg|_{\alpha=0}
        \right)\alpha\\
        \label{modyfyingNotation2}
        &=
        \sum_{i=1}^{n}
        \left(
        \frac{\partial G[\{G^j_{\gamma}(\alpha|i)(\circ): j=1,\ldots,m\}](x)}{\partial \alpha}\Bigg|_{\alpha=0}
        \right)\alpha\\
        \label{usingChainRuleForAlpha}
        &=
        \sum_{i=1}^{n}\sum_{j=1}^{m}
        \left(
        \frac{\partial G[G^1_{\gamma}(0|i)(\circ),\ldots,G^j_{\gamma}(\alpha|i)(\circ),\ldots,G^m_{\gamma}(0|i)(\circ)](x)}{\partial \alpha}\Bigg|_{\alpha=0}
        \right)\alpha
    \end{align}
\end{subequations}
in \cref{expandingDeltaGInAlpha} we have performed a series expansion in infinitesimal real parameter $\alpha$, while in \cref{modyfyingNotation2} we have used the notation $G^j_{\gamma}(\alpha|i)(\circ)\equiv G^j[h_1(\cdot),\ldots,h_i(\cdot)+\alpha \gamma_i(\cdot),\ldots,h_n(\cdot)](\circ)$. In \cref{usingChainRuleForAlpha} we have applied the ordinary chain rule for differentiation with respect to $\alpha$. We may treat each $G^j_{\gamma}(\alpha|i)$ as a function of the variable $\alpha$, and expand around $\alpha=0$ to first order as follows:
    \begin{align}
        \label{ExpandingGjinAlpha}
        \delta G(x)
        &=
        \sum_{i=1}^{n}
        \sum_{j=1}^{m}
        \frac{\partial}{\partial \alpha}
        \bigg(
            G
            \big[
                G^1_{\gamma}(0|i)(\circ),\ldots,
                G^j_{\gamma}(0|i)(\circ)
                +
                \alpha
                \left(
                \frac{\partial G^j_\gamma(\alpha|i)(\circ)}{\partial \alpha}\Bigg|_{\alpha=0}
                \right),
                \ldots,
                G^m_{\gamma}(0|i)(\circ)
            \big](x)
        \bigg)
        \Bigg|_{\alpha=0}
        \alpha
    \end{align}
Now, recall the definition of the functional derivative in \cref{functionalDerivativeDefinition}, which we restate here for an arbitrary function-valued functional $\tilde{G}$ for the sake of the reader's convenience:
\begin{equation}
    \label{functionalDerivativeDefinition3}
    \int d^d z \frac{\partial \tilde{G}(x)}{\partial h_j(z)} \gamma_j(z) \equiv \frac{\partial \tilde{G}[h_1(\cdot),\ldots,h_j(\cdot)+\alpha \gamma_j(\cdot),\ldots,h_n(\cdot)](x)}{\partial \alpha}\Bigg|_{\alpha=0},
\end{equation}
Comparing \cref{ExpandingGjinAlpha} with \cref{functionalDerivativeDefinition3}, we see that we may write:
\begin{subequations}
    \begin{align}
        \delta G(x)
        &=
        \left(
            \sum_{i=1}^{n}
            \sum_{j=1}^{m}
            \int d^d z~
            \frac{\partial G(x)}{\partial G^j(0|j)(z)}
            \frac{\partial G^j(\alpha|i)(z)}{\partial \alpha}\Bigg|_{\alpha=0}
        \right)
        \alpha
        \\
        \label{undoingNotaion}
        &=
        \sum_{j=1}^{m}
        \int d^d z~
        \frac{\partial G(x)}{\partial G^j(z)}
        \left(
            \sum_{i=1}^{n}
            \frac{\partial G^j(\alpha|i)(z)}{\partial \alpha}\Bigg|_{\alpha=0}\alpha
        \right)\\
        \label{usingVariationDefinition}
        &=
        \sum_{j=1}^{m}
        \int d^d z~
        \frac{\partial G(x)}{\partial G^j(z)}
        ~\delta G^j(z),
    \end{align}
\end{subequations}
in \cref{undoingNotaion} we have undone the $G^j(\alpha|i)(z)$ notation introduced earlier, while in \cref{usingVariationDefinition} we have used the definition of the variation from \cref{variationDefinition}. Thus, we have shown that:
\begin{equation}
    \delta G(x)
    =
    \sum_{j=1}^{m}
    \int d^d z~
    \frac{\partial G(x)}{\partial G^j(z)}
    ~\delta G^j(z),
\end{equation}
as stated in \cref{variationChainRule}.

\subsection{Proof of the Chain Rule for Spacetime Derivatives}
\label{DerivativeChainRuleProof}
Here we shall prove the chain rule for spacetime derivatives in \cref{derivativeChainRule}. We assume that the function-valued functional $G\in \mathscr{G}$ has the following translational property:
\begin{subequations}
    \label{TranslationalPropertyOfGRestated}
    \begin{align}
        G[h_1(\cdot+y),\ldots,h_n(\cdot+y)](x)
        &=
        G[h_1(\cdot),\ldots,h_n(\cdot)](x+y).
    \end{align}
\end{subequations}
Suppose we are now to compose $G$ with a set of function-valued functionals $\{G^j\in  \mathscr{G}: j=1,\ldots,m\}$, that is
\begin{subequations}
    \begin{align}
        G(x)=G[\{G^j[h_1(\cdot),\ldots,h_n(\cdot)](\circ): j=1,\ldots,m\}](x).
    \end{align}
\end{subequations}
To continue with the proof, we shall define the following useful notation:
\begin{subequations}
    \begin{align}
        \label{GAlphaDefinition}
        G^j_{\alpha}(x)
        &\equiv
        G^j[h_1(\cdot),\ldots,h_n(\cdot)](x+\alpha e^\nu) \\
        \label{gammaAlphaDefinition}
        \gamma^j_{\alpha}(x)
        &\equiv
        \frac{G_\alpha^j(x)-G^j_0(x)}{\alpha},
    \end{align}
\end{subequations}
where $e^\nu$ is a unit vector in the $\nu$ direction and $G^j_0(x)=G^j[h_1(\cdot),\ldots,h_n(\cdot)](x)$. Now, to compute the spacetime derivative of $G$, we begin with the definition of a spacetime derivative in the $\nu$ direction:
\begin{subequations}
    \begin{align}
        \partial_\nu^x G(x)
        &\equiv
        \underset{\alpha\to0}{\lim}
        \frac{G(x+\alpha e^\nu)-G(x)}{\alpha}\\
        &=
        \underset{\alpha\to0}{\lim}
        \frac{1}{\alpha}
        \bigg(
              G[\{G^j[h_1(\cdot),\ldots,h_n(\cdot)](\circ): j=1,\ldots,m\}](x+\alpha e^\nu)\\
              \notag
              &\qquad \quad ~ ~ -
              G[\{G^j[h_1(\cdot),\ldots,h_n(\cdot)](\circ): j=1,\ldots,m\}](x)  
        \bigg)\\
        \label{usingTranslationalPropertyOfG}
        &=
        \underset{\alpha\to0}{\lim}
        \frac{1}{\alpha}
        \bigg(
              G[\{G^j[h_1(\cdot),\ldots,h_n(\cdot)](\circ+\alpha e^\nu): j=1,\ldots,m\}](x)\\
              \notag
              &\qquad \quad ~ ~ -
              G[\{G^j[h_1(\cdot),\ldots,h_n(\cdot)](\circ): j=1,\ldots,m\}](x)  
        \bigg)\\
        \label{usingCondensedNotation}
        &=
        \underset{\alpha\to0}{\lim}
        \frac{1}{\alpha}
        \bigg(
              G[\{G^j_\alpha(\circ): j=1,\ldots,m\}](x)
              -
              G[\{G^j_0(\circ): j=1,\ldots,m\}](x)  
        \bigg),
    \end{align}
\end{subequations}
in \cref{usingTranslationalPropertyOfG} we have used the translational property of $G$ from \cref{TranslationalPropertyOfGRestated}, while in \cref{usingCondensedNotation} we have used the condensed notation introduced in \cref{GAlphaDefinition}. Now, we may treat $G$ as a function of the variable $\alpha$, and expand around $\alpha=0$ to first order as follows:
\begin{subequations}
    \begin{align}
        \partial_\nu^x G(x)
        &=
        \underset{\alpha\to0}{\lim}
        \frac{1}{\alpha}
        \bigg(
        \frac{\partial G[\{G^j_\alpha(\circ): j=1,\ldots,m\}](x)}{\partial \alpha}\Bigg|_{\alpha=0}
        \alpha
        \bigg)\\
        \label{usingGammaAlphaDefinition}
        &=
        \frac{\partial G[\{G_0^j(\circ)+\alpha \gamma^j_\alpha(\circ): j=1,\ldots,m\}](x)}{\partial \alpha}\Bigg|_{\alpha=0}\\
        \label{applyingOrdinaryChainRule}
        &=
        \sum_{j=1}^{m}
        \frac
        {\partial G[G_0^1(\circ),\ldots,G_0^j(\circ)+\alpha \gamma^j_\alpha(\circ),\ldots,G_0^n(\circ)](x)}
        {\partial \alpha}\Bigg|_{\alpha=0},
    \end{align}
\end{subequations}
in \cref{usingGammaAlphaDefinition} we have used the definition of $\gamma^j_\alpha$ from \cref{gammaAlphaDefinition} to write $G^j_\alpha(\circ)=G_0^j(\circ)+\alpha \gamma^j_\alpha(\circ)$, and in \cref{applyingOrdinaryChainRule} we have applied the ordinary chain rule for differentiation with respect to real variable $\alpha$. Recall the definition of the functional derivative in \cref{functionalDerivativeDefinition}, which we restate here for an arbitrary function-valued functional $\tilde{G}$ for the sake of the reader's convenience:
\begin{equation}
    \label{functionalDerivativeDefinition4}
    \int d^d z \frac{\partial \tilde{G}(x)}{\partial h_j(z)} \gamma_j(z) \equiv \frac{\partial \tilde{G}[h_1(\cdot),\ldots,h_j(\cdot)+\alpha \gamma_j(\cdot),\ldots,h_n(\cdot)](x)}{\partial \alpha}\Bigg|_{\alpha=0},
\end{equation}
Comparing \cref{applyingOrdinaryChainRule} with \cref{functionalDerivativeDefinition4}, we see that we may write:
\begin{subequations}
    \begin{align}
        \partial_\nu^x G(x)
        &=
        \sum_{j=1}^{m}
        \int d^d z~
        \frac{\partial G(x)}{\partial G_0^j(z)}
        \bigg(
        \underset{\alpha\to0}{\lim}~
        \gamma^j_\alpha(z)
        \bigg)\\
        \label{undoingAlphaNotation}
        &=
        \sum_{j=1}^{m}
        \int d^d z~
        \frac{\partial G(x)}{\partial G^j(z)}
        ~\left(
            \underset{\alpha\to0}{\lim}~
            \frac{G^j[h_1(\cdot),\ldots,h_n(\cdot)](z+\alpha e^\nu)-G^j[h_1(\cdot),\ldots,h_n(\cdot)](z)}{\alpha}
        \right)\\
        \label{finalStepOfDerivativeChainRuleProof}
        &=
        \sum_{j=1}^{m}
        \int d^d z~
        \frac{\partial G(x)}{\partial G^j(z)}
        ~\partial_\nu^z G^j(z),
    \end{align}
\end{subequations}
in \cref{undoingAlphaNotation} we have expanded the $G^j_\alpha(z)$ and $\gamma^j_\alpha$ notation introduced in \cref{GAlphaDefinition,gammaAlphaDefinition}, while in \cref{finalStepOfDerivativeChainRuleProof} we have used the definition of the spacetime derivative. Thus, we have shown that:
\begin{equation}
    \partial_\nu^x G(x)
    =
    \sum_{j=1}^{m}
    \int d^d z~
    \frac{\partial G(x)}{\partial G^j(z)}
    ~\partial_\nu^z G^j(z),
\end{equation}
as stated in \cref{derivativeChainRule}.

\subsection{Proof of Product Rule for Variations}
\label{ProductRuleForVariationsProof}
In this section, we shall prove the product rule for variations. Consider two function-valued functionals $G_1,G_2\in \mathscr{G}$, with $G_1$ and $G_2$ dependent on $n_1$ and $n_2$ functions respectively. Now, take $n=\max(n_1,n_2)$, so that we may write $G_1(x)=G_1[h_1(\cdot),\ldots,h_n(\cdot)](x)$ and $G_2(x)=G_2[h_1(\cdot),\ldots,h_n(\cdot)](x)$. We may then form the function-valued functional $G_3\in \mathscr{G}$ defined as the product of $G_1$ and $G_2$:
\begin{equation}
    G_3(x)\equiv G_1(x)G_2(x),
\end{equation}
and we may write
\begin{equation}
    G_3(x)=G_3[h_1(\cdot),\ldots,h_n(\cdot)](x)=G_1[h_1(\cdot),\ldots,h_n(\cdot)](x)G_2[h_1(\cdot),\ldots,h_n(\cdot)](x).
\end{equation}
Now, making use of \cref{variationDefinitionExpandedITOderivative}, we compute the variation of $G_3$ as:
\begin{subequations}
    \begin{align}
        \delta G_3(x)
        &=
        \sum_{i=1}^{n}
        \left(
            \frac{\partial G_3[h_1(\cdot),\ldots,h_i(\cdot)+\alpha \gamma_i(\cdot),\ldots,h_n(\cdot)](x)}{\partial \alpha}\Bigg|_{\alpha=0}
        \right)\alpha\\
        &=
        \sum_{i=1}^{n}
        \left(
            \frac{\partial
            \big(
                G_1[h_1(\cdot),\ldots,h_i(\cdot)+\alpha \gamma_i(\cdot),\ldots,h_n(\cdot)](x)
                ~
                G_2[h_1(\cdot),\ldots,h_i(\cdot)+\alpha \gamma_i(\cdot),\ldots,h_n(\cdot)](x)
            \big)}{\partial \alpha}\Bigg|_{\alpha=0}
        \right)\alpha\\
        \label{applyingOrdinaryProductRule}
        &=
        \sum_{i=1}^{n}
        \bigg(
            \frac{\partial
            G_1[h_1(\cdot),\ldots,h_i(\cdot)+\alpha \gamma_i(\cdot),\ldots,h_n(\cdot)](x)}{\partial \alpha}\Bigg|_{\alpha=0}
            ~
            G_2[h_1(\cdot),\ldots,h_n(\cdot)](x)\\
            \notag
            &~~~+
            G_1[h_1(\cdot),\ldots,h_n(\cdot)](x)
            ~
            \frac{\partial
            G_2[h_1(\cdot),\ldots,h_i(\cdot)+\alpha \gamma_i(\cdot),\ldots,h_n(\cdot)](x)}{\partial \alpha}\Bigg|_{\alpha=0}
        \bigg)\alpha
    \end{align}
\end{subequations}
in \cref{applyingOrdinaryProductRule} we have applied the ordinary product rule for differentiation with respect to $\alpha$. Now, using the definition of the variation from \cref{variationDefinition}, we may write:
\begin{equation}
    \delta G_3(x)
    =
    \delta G_1(x) ~ G_2(x)
    +
    G_1(x) ~ \delta G_2(x),
\end{equation}
thus proving the product rule for variations.

\section{Useful Identities for the Proof of the Transposition Rule}
\subsection{Translational Property of \texorpdfstring{$B$}{B}}
\label{translationalPropertyOfB}
The function-valued functional $B$ defined in \cref{definitionOfBFunctional} has the form
\begin{equation}
    \label{definitionOfBFunctional2}
    B=B[\partial_\nu A^\mu(\cdot): \nu,\mu\neq\mu_D](x)
    \equiv
    -\sum_{\nu\neq\mu_D} \partial_\nu f^\nu [A^\mu:\mu\neq\mu_D](x).
\end{equation}
Recall that the function-valued functional $f^\nu$ was assumed to have the functional power-series expansion in \cref{fNuAsPowerSeriesOfIndependentFieldComponents}. Thus, we may write $B$ as
\begin{equation}
    B[\partial_\nu A^\mu(\cdot): \nu,\mu\neq\mu_D](x)
    =
    -\sum_{\nu\neq\mu_D} \partial_\nu \left(
    \sum_{I=0}^{\infty} 
    c^\nu_{i_0,\ldots,i_{d-1}}
    (A^0(x))^{i_0} \ldots(A^{\mu_D-1}(x))^{i_{\mu_D-1}}(A^{\mu_D+1}(x))^{i_{\mu_D+1}} \ldots (A^{d-1}(x))^{i_{d-1}}
    \right).
\end{equation}
Now, if one considers $B[\partial_\nu A^\mu(\cdot): \nu,\mu\neq\mu_D](x+y)$, then one will find:
\begin{subequations}
    \begin{align}
        B[\partial_\nu A^\mu(\cdot): \nu,\mu\neq\mu_D](x+y)
        &=
        -\sum_{\nu\neq\mu_D} \partial_\nu^x \bigg(
        \sum_{I=0}^{\infty} 
        c^\nu_{i_0,\ldots,i_{d-1}}
        \\
        \notag
        &\qquad\qquad\qquad
        (A^0(x+y))^{i_0} \ldots(A^{\mu_D-1}(x+y))^{i_{\mu_D-1}}(A^{\mu_D+1}(x+y))^{i_{\mu_D+1}} \ldots (A^{d-1}(x+y))^{i_{d-1}}
        \bigg)\\
        &=
        B[\partial_\nu A^\mu(\cdot+y): \nu,\mu\neq\mu_D](x).
    \end{align}
\end{subequations}
Thus, we have shown that $B[\partial_\nu A^\mu(\cdot): \nu,\mu\neq\mu_D](x+y)=B[\partial_\nu A^\mu(\cdot+y): \nu,\mu\neq\mu_D](x)$, which will allow us to make use of the chain rule for spacetime derivatives from \cref{derivativeChainRule} when we compute the spacetime derivative of $B$. That is, we may write
\begin{equation}
    \label{spaceTimeDerivativeOfB}
    \partial^x_\mu B[\partial_\nu A^\mu(\cdot): \nu,\mu\neq\mu_D](x)
    =
    \sum_{\rho,\sigma\neq\mu_D} \int d^d z~
    \frac{\partial B(x)}{\partial (\partial^z_\rho A^\sigma(z))}
    ~\partial_\mu^z \big(\partial^z_\rho A^\sigma(z)\big).
\end{equation}

\subsection{\texorpdfstring{$2^{nd}$}{2nd} Order Functional Derivatives Commute}
Using \cref{gammaAsDeltaFunction}, we may write the second order functional derivative of a functional $F\in \mathscr{F}$ as
\begin{equation}
    \label{secondOrderFunctionalDerivativeAsAlphaDerivative}
    \frac{\partial^2 F}{\partial h_j(z)\partial h_i(y)} 
    = 
    \frac
    {\partial^2 F[h_1(\cdot),\ldots,h_i(\cdot)+\alpha \delta^{d}(\cdot - y),\ldots,h_j(\cdot)+\beta \delta^{d}(\cdot - z),\ldots,h_n(\cdot)]}
    {\partial \beta \partial \alpha}\Bigg|_{\alpha=\beta=0},
\end{equation}
where we assume that the partial derivatives $\partial_\alpha F$ and $\partial_\beta F$ exist and are themselves differentiable with respect to both $\alpha$ and $\beta$. Thus, the mixed partial derivatives $\partial^2 F/\partial \alpha \partial \beta$ and $\partial^2 F/\partial \beta \partial \alpha$ also exist and are equal by Schwarz's theorem \cite{apostol1974mathematical,rudin1976principles}. Therefore, we have
\begin{equation}
    \label{secondOrderFunctionalDerivativeAsAlphaDerivativeCommuted}
    \frac{\partial^2 F}{\partial h_j(z)\partial h_i(y)} 
    = 
    \frac{\partial^2 F}{\partial h_i(y)\partial h_j(z)}.
\end{equation}

\subsection{Variation of the Functional Derivative of \texorpdfstring{$B$}{B}}
The functional derivative of $B$ with respect to $\partial^y_\rho A^\sigma(y)$ is given by
\begin{equation}
    \label{variationOfFunctionalDerivativeOfB1}
    \frac{\partial B(x)}{\partial (\partial^y_\rho A^\sigma(y))}
    =
    \frac
    {\partial B[\ldots,\partial_\rho A^\sigma(\cdot)+\alpha \delta^d(\cdot-y),\ldots](x)}
    {\partial \alpha}\Bigg|_{\alpha=0}.
\end{equation}
Taking the variation of this expression, we have
\begin{subequations}
    \begin{align}
        \delta\left(\frac{\partial B(x)}{\partial (\partial^y_\rho A^\sigma(y))}\right)
        &=
        \delta\left(
        \frac
        {\partial B[\ldots,\partial_\rho A^\sigma(\cdot)+\alpha \delta^d(\cdot-y),\ldots](x)}
        {\partial \alpha}\Bigg|_{\alpha=0}
        \right)\\
        &=
        \sum_{\rho^\prime,\sigma^\prime\neq\mu_D}
        \frac{\partial}{\partial \beta}
        \left(
            \frac
            {
                \partial B
                [
                    \ldots,
                    \partial_\rho A^\sigma(\cdot)+\alpha \delta^d(\cdot-y),
                    \ldots,
                    \partial_{\rho^\prime}A^{\sigma^\prime}(\cdot)+\beta \gamma^{\sigma^\prime}_{\rho^\prime},
                    \ldots
                ](x)
            }
            {\partial \alpha}\Bigg|_{\alpha=0}
        \right)\Bigg|_{\beta=0}
        \beta\\
        &=
        \frac{\partial}{\partial \alpha}
        \left(
            \sum_{\rho^\prime,\sigma^\prime\neq\mu_D}
            \frac
            {
                \partial B
                [
                    \ldots,
                    \partial_\rho A^\sigma(\cdot)+\alpha \delta^d(\cdot-y),
                    \ldots,
                    \partial_{\rho^\prime}A^{\sigma^\prime}(\cdot)+\beta \gamma^{\sigma^\prime}_{\rho^\prime},
                    \ldots
                ](x)
            }
            {\partial \beta}\Bigg|_{\beta=0}\beta
        \right)\Bigg|_{\alpha=0}\\
        \label{defineAsigmaalpha}
        &=
        \frac{\partial}{\partial \alpha}
        \left(
            \delta B
                [
                    \ldots,
                    \partial_\rho A^{\sigma,\alpha}_y(\cdot),
                    \ldots
                ](x)
        \right)\Bigg|_{\alpha=0}.
    \end{align}
\end{subequations}
In \cref{defineAsigmaalpha} we have defined $\partial_\rho A^{\sigma,\alpha}_y(\cdot)\equiv \partial_\rho A^\sigma(\cdot)+\alpha \delta^d(\cdot-y)$. Next, we will use the chain rule for variations from \cref{variationChainRule} to write $\delta B$ in terms of the variations of its arguments as follows:
\begin{align}
    \label{variationOfBBrokenSum}
        \delta B(x)
        &=
        \int d^dz~ 
        \frac
        {\partial B[\ldots,\partial_\rho A^{\sigma,\alpha}_y(\cdot),\ldots](x)}
        {\partial (\partial_\rho^z A^{\sigma,\alpha}_y(z))}
        ~\delta(\partial_\rho^z A^{\sigma,\alpha}_y(z))\\
        \notag
        &+
        \sum_{(\rho^\prime,\sigma^\prime)\neq(\rho,\sigma,\mu_D)}
        \int d^dz~ 
        \frac
        {\partial B[\ldots,\partial_\rho A^{\sigma,\alpha}_y(\cdot),\ldots](x)}
        {\partial (\partial_{\rho^\prime}^z A^{\sigma^\prime}(z))}
        ~\delta(\partial_{\rho^\prime}^z A^{\sigma^\prime}(z)),
\end{align}
where in \cref{variationOfBBrokenSum} we have broken the sum over $\rho^\prime$ and $\sigma^\prime$ into the term where $(\rho^\prime,\sigma^\prime)=(\rho,\sigma)$, and the terms where $(\rho^\prime,\sigma^\prime)\neq(\rho,\sigma)$. Inserting \cref{variationOfBBrokenSum} into \cref{defineAsigmaalpha}, we have
\begin{subequations}
    \begin{align}
        \delta\left(\frac{\partial B(x)}{\partial (\partial^y_\rho A^\sigma(y))}\right)
        &=
        \int d^dz~ 
        \frac
        {\partial B[\ldots,\partial_\rho A^{\sigma}(\cdot),\ldots](x)}
        {\partial (\partial_\rho^z A^{\sigma}(z))}
        ~\frac
        {\partial}{\partial \alpha}
        \left(
        \delta(\partial_\rho^z A^{\sigma,\alpha}_y(z))
        \right)\Bigg|_{\alpha=0}\\
        \notag
        &+
        \int d^dz~ 
        \frac
        {\partial}{\partial \alpha}
        \left(
        \frac
        {\partial B[\ldots,\partial_\rho A^{\sigma,\alpha}_y(\cdot),\ldots](x)}
        {\partial (\partial_\rho^z A^{\sigma,\alpha}_y(z))}
        \right)\Bigg|_{\alpha=0}
        ~\delta(\partial_\rho^z A^{\sigma}(z))\\
        \notag
        &+
        \sum_{(\rho^\prime,\sigma^\prime)\neq(\rho,\sigma,\mu_D)}
        \int d^dz~ 
        \frac
        {\partial}{\partial \alpha}
        \left(
        \frac
        {\partial B[\ldots,\partial_\rho A^{\sigma,\alpha}_y(\cdot),\ldots](x)}
        {\partial (\partial_{\rho^\prime}^z A^{\sigma^\prime}(z))}
        \right)\Bigg|_{\alpha=0}
        ~\delta(\partial_{\rho^\prime}^z A^{\sigma^\prime}(z))\\
        \label{recombineSums}
        &=
        \int d^dz~ 
        \frac
        {\partial B[\ldots,\partial_\rho A^{\sigma}(\cdot),\ldots](x)}
        {\partial (\partial_\rho^z A^{\sigma}(z))}
        ~\frac
        {\partial}{\partial \alpha}
        \left(
        \delta(\partial_\rho^z A^{\sigma,\alpha}_y(z))
        \right)\Bigg|_{\alpha=0}\\
        \notag
        &+
        \sum_{(\rho^\prime,\sigma^\prime)\neq(\mu_D)}
        \int d^dz~ 
        \frac
        {\partial}{\partial \alpha}
        \left(
        \frac
        {\partial B[\ldots,\partial_\rho A^{\sigma,\alpha}_y(\cdot),\ldots](x)}
        {\partial (\partial_{\rho^\prime}^z A^{\sigma^\prime}(z))}
        \right)\Bigg|_{\alpha=0}
        ~\delta(\partial_{\rho^\prime}^z A^{\sigma^\prime}(z))\\
        &=
        \label{recombinedSums2}
        \int d^dz~ 
        \frac
        {\partial B[\ldots,\partial_\rho A^{\sigma}(\cdot),\ldots](x)}
        {\partial (\partial_\rho^z A^{\sigma}(z))}
        ~\frac
        {\partial}{\partial \alpha}
        \left(
        \delta(\partial_\rho^z A^{\sigma,\alpha}_y(z))
        \right)\Bigg|_{\alpha=0}\\
        \notag
        &+
        \sum_{(\rho^\prime,\sigma^\prime)\neq(\mu_D)}
        \int d^dz~ 
        \frac
        {\partial^2 B(x)}
        {
        \partial
        (\partial_\rho^z A^{\sigma}(y)) 
        \partial
        (\partial_{\rho^\prime}^z A^{\sigma^\prime}(z))}
        ~\delta(\partial_{\rho^\prime}^z A^{\sigma^\prime}(z))
    \end{align}
\end{subequations}
in \cref{recombineSums} we have recombined the sums over $\rho^\prime$ and $\sigma^\prime$ back into a single sum. Then, if one pays attention to the remaining $\alpha$ derivative in \cref{recombinedSums2}, one will find:
\begin{subequations}
    \begin{align}
        \label{useTranspositionRuleForIndependentFieldComponents}
        ~\frac
        {\partial}{\partial \alpha}
        \left(
        \delta(\partial_\rho^z A^{\sigma,\alpha}_y(z))
        \right)\Bigg|_{\alpha=0}
        &=
        ~\frac
        {\partial}{\partial \alpha}
        \left(
        \partial_\rho(\delta A^{\sigma,\alpha}(z))
        \right)\Bigg|_{\alpha=0}\\
        &=
        ~\frac
        {\partial}{\partial \alpha}
        \left(
        \partial_\rho
        ( 
            A^{\sigma,\alpha}(z)+\beta \gamma^\sigma_\rho(z)-A^{\sigma,\alpha}(z)
        )
        \right)\Bigg|_{\alpha=0}\\
        \label{useIndependeceOfAlpha}
        &=
        \partial_\rho
        ~\frac
        {\partial}{\partial \alpha}
        \left(
            \beta \gamma^\sigma_\rho(z)
        \right)\Bigg|_{\alpha=0}\\
        &=0,
    \end{align}
\end{subequations}
where in \cref{useTranspositionRuleForIndependentFieldComponents} we have the transposition rule for independent field components from \cref{TranspositionRuleIndependentComponent}, and in \cref{useIndependeceOfAlpha} we have used the fact that $\alpha$ and $\beta$ are independent variables, so that $\partial_\alpha \beta=0$. Thus, we have shown that
\begin{subequations}
    \begin{align}
    \delta\left(\frac{\partial B(x)}{\partial (\partial^y_\rho A^\sigma(y))}\right)
        &=
        \sum_{(\rho^\prime,\sigma^\prime)\neq(\mu_D)}
        \int d^dz~ 
        \frac
        {\partial^2 B(x)}
        {
        \partial
        (\partial_\rho^y A^{\sigma}(y)) 
        \partial
        (\partial_{\rho^\prime}^z A^{\sigma^\prime}(z))}
        ~\delta(\partial_{\rho^\prime}^z A^{\sigma^\prime}(z))\\
        \label{variationOfFunctionalDerivativeOfBFinal}
        &=        
        \sum_{(\rho^\prime,\sigma^\prime)\neq(\mu_D)}
        \int d^dz~ 
        \frac
        {\partial^2 B(x)}
        {\partial(\partial_{\rho^\prime}^z A^{\sigma^\prime}(z)) \partial(\partial_\rho^y A^{\sigma}(y)) }
        ~\delta(\partial_{\rho^\prime}^z A^{\sigma^\prime}(z)),
    \end{align}
\end{subequations}
in \cref{variationOfFunctionalDerivativeOfBFinal} we commuted the order of the functional derivatives using the result from \cref{secondOrderFunctionalDerivativeAsAlphaDerivativeCommuted}.

\subsection{\texorpdfstring{$2^{nd}$}{2nd} Order Functional Derivative of \texorpdfstring{$B$}{B}}
Consider acting the spacetime derivative $\partial_\mu^x$ on the functional derivative of $B$ with respect to $\partial_\rho^z A^\sigma(z)$:
\begin{subequations}
    \begin{align}
        \partial_\mu^x 
        \left(
            \frac{\partial B(x)}{\partial (\partial^z_\rho A^\sigma(z))}
        \right)
        &=
        \partial_\mu^x 
        \left(
            \frac
            {\partial B[\ldots,\partial_\rho A^{\sigma}(\cdot)+\alpha\delta^d(\cdot-z),\ldots](x)}
            {\partial \alpha}
            \bigg|_{\alpha=0}
        \right)
        \\
        &=
         \frac
            {\partial }
            {\partial \alpha} 
        \left(
           \partial_\mu^x
           B[\ldots,\partial_\rho A^{\sigma}(\cdot)+\alpha\delta^d(\cdot-z),\ldots](x) 
        \right)\bigg|_{\alpha=0}
        \\
        \label{defineAsigmaAlphaAgain}
        &=
         \frac
            {\partial }
            {\partial \alpha} 
        \left(
           \partial_\mu^x
           B[\ldots,\partial_\rho A^{\sigma,\alpha}_z(\cdot),\ldots](x) 
        \right)\bigg|_{\alpha=0},
    \end{align}
\end{subequations}
in \cref{defineAsigmaAlphaAgain} we have defined $\partial_\rho A^{\sigma,\alpha}_z(\cdot)\equiv \partial_\rho A^\sigma(\cdot)+\alpha \delta^d(\cdot-z)$. In appendix \ref{translationalPropertyOfB}, we showed that $B$ has the translational property, which allows us to make use of the chain rule for spacetime derivatives from \cref{derivativeChainRule} to write $\partial_\mu^x B$ in terms of the functional derivatives of $B$. Thus, we have
\begin{subequations}
    \begin{align}
        \label{breakUpSumInSpaceTimeChainRule}
        \partial_\mu^x 
        \left(
            \frac{\partial B(x)}{\partial (\partial^z_\rho A^\sigma(z))}
        \right)
        &=
        \frac{\partial}{\partial \alpha}
        \bigg(
        \int d^dy~
        \frac
        {\partial B[\ldots,\partial_\rho A^{\sigma,\alpha}_z(\cdot),\ldots](x)}
        {\partial (\partial^y_\rho A^{\sigma,\alpha}_z(y))}
        \partial_\mu^y \big(\partial^y_\rho A^{\sigma,\alpha}_z(y)\big)
        \\
        \notag
        &+
        \sum_{(\rho^\prime,\sigma^\prime)\neq(\rho,\sigma,\mu_D)}
        \int d^dy~
        \frac
        {\partial B[\ldots,\partial_\rho A^{\sigma,\alpha}_z(\cdot),\ldots](x)}
        {\partial (\partial^y_{\rho^\prime} A^{\sigma^\prime}(y))}
        \partial_\mu^y \big(\partial^y_{\rho^\prime} A^{\sigma^\prime}_z(y)\big)
        \bigg)\bigg|_{\alpha=0}
        \\
        &=
        \int d^dy~
        \frac
        {\partial B[\ldots,\partial_\rho A^{\sigma}(\cdot),\ldots](x)}
        {\partial (\partial^y_\rho A^{\sigma}(y))}
        \frac{\partial}{\partial \alpha}
        \left(
        \partial_\mu^y \big(\partial^y_\rho A^{\sigma,\alpha}_z(y)\big)
        \right)\bigg|_{\alpha=0}\\
        \notag
        &+
        \int d^dy~
        \frac{\partial}{\partial \alpha}
        \left(
        \frac
        {\partial B[\ldots,\partial_\rho A^{\sigma,\alpha}_z(\cdot),\ldots](x)}
        {\partial (\partial^y_\rho A^{\sigma,\alpha}_z(y))}
        \right)\bigg|_{\alpha=0}
        \partial_\mu^y \big(\partial^y_\rho A^{\sigma}(y)\big)\\
        \notag
        &+
        \sum_{(\rho^\prime,\sigma^\prime)\neq(\rho,\sigma,\mu_D)}
        \int d^dy~
        \frac{\partial}{\partial \alpha}
        \left(
        \frac
        {\partial B[\ldots,\partial_\rho A^{\sigma,\alpha}_z(\cdot),\ldots](x)}
        {\partial (\partial^y_{\rho^\prime} A^{\sigma^\prime}(y))}
        \right)\bigg|_{\alpha=0}
        \partial_\mu^y \big(\partial^y_{\rho^\prime} A^{\sigma^\prime}_z(y)\big)\\
        &=
        \label{alphaDerivativeOfSpaceTimeChainRule}
        \int d^dy~
        \frac
        {\partial B[\ldots,\partial_\rho A^{\sigma}(\cdot),\ldots](x)}
        {\partial (\partial^y_\rho A^{\sigma}(y))}
        \frac{\partial}{\partial \alpha}
        \left(
        \partial_\mu^y \big(\partial^y_\rho A^{\sigma,\alpha}_z(y)\big)
        \right)\bigg|_{\alpha=0}\\
        \notag
        &+
        \sum_{(\rho^\prime,\sigma^\prime)\neq(\mu_D)}
        \int d^dy~
        \frac
        {\partial^2 B(x)}
        {\partial (\partial^z_\rho A^{\sigma}(z)) \partial (\partial^y_{\rho^\prime} A^{\sigma^\prime}(y))}
        \partial_\mu^y \big(\partial^y_{\rho^\prime} A^{\sigma^\prime}_z(y)\big),
    \end{align}
\end{subequations}
in \cref{breakUpSumInSpaceTimeChainRule} we have broken the sum from the spacetime chain rule into two pieces, one that runs over all $(\rho^\prime,\sigma^\prime)\neq(\rho,\sigma,\mu_D)$, and then we included the remaining $\rho,\sigma$ part of the sum.  If we now shift our focus to the remaining $\alpha$ derivative in \cref{alphaDerivativeOfSpaceTimeChainRule}, we will find the following:
    \begin{align}
        \label{remainingAlphaDerivativeEvaluated}
        \frac{\partial}{\partial \alpha}
        \left(
        \partial_\mu^y \big(\partial^y_\rho A^{\sigma,\alpha}_z(y)\big)
        \right)\bigg|_{\alpha=0}
        =
        \frac{\partial}{\partial \alpha}
        \left(
        \partial_\mu^y 
        \big[
            \partial^y_\rho A^{\sigma}(y)+\alpha\delta^d(y-z)
        \big]
        \right)\bigg|_{\alpha=0}
        =
        \partial^y_\mu \delta^d(y-z).
    \end{align}
Inserting \cref{remainingAlphaDerivativeEvaluated} into \cref{alphaDerivativeOfSpaceTimeChainRule}, we find
\begin{subequations}
    \begin{align}
        \partial_\mu^x 
        \left(
            \frac{\partial B(x)}{\partial (\partial^z_\rho A^\sigma(z))}
        \right)
        &=
        \int d^dy~
        \frac
        {\partial B[\ldots,\partial_\rho A^{\sigma}(\cdot),\ldots](x)}
        {\partial (\partial^y_\rho A^{\sigma}(y))}
        \partial^y_\mu \delta^d(y-z)
        \\
        \notag
        &+
        \sum_{(\rho^\prime,\sigma^\prime)\neq(\mu_D)}
        \int d^dy~
        \frac
        {\partial^2 B(x)}
        {\partial (\partial^z_\rho A^{\sigma}(z)) \partial (\partial^y_{\rho^\prime} A^{\sigma^\prime}(y))}
        \partial_\mu^y \big(\partial^y_{\rho^\prime} A^{\sigma^\prime}_z(y)\big)
        \\
        &=
        \label{useIBP}
        -
        \partial^z_\mu
        \left(
        \frac
        {\partial B(x)}
        {\partial (\partial^y_\rho A^{\sigma}(y))}
        \right) 
        \\
        \notag
        &+
        \sum_{(\rho^\prime,\sigma^\prime)\neq(\mu_D)}
        \int d^dy~
        \frac
        {\partial^2 B(x)}
        {\partial (\partial^z_\rho A^{\sigma}(z)) \partial (\partial^y_{\rho^\prime} A^{\sigma^\prime}(y))}
        \partial_\mu^y \big(\partial^y_{\rho^\prime} A^{\sigma^\prime}_z(y)\big),
    \end{align}
\end{subequations}
where in \cref{useIBP}, we used integration by parts to move the spacetime derivative off the $\delta$. Finally, we may rearrange \cref{useIBP} to obtain
\begin{equation}
    \label{secondOrderDerivativeOfBFinal2}
    \partial_\mu^x 
        \left(
            \frac{\partial B(x)}{\partial (\partial^z_\rho A^\sigma(z))}
        \right)
    +
    \partial^z_\mu
        \left(
            \frac{\partial B(x)}{\partial (\partial^y_\rho A^{\sigma}(y))}
        \right)
    =
    \sum_{(\rho^\prime,\sigma^\prime)\neq(\mu_D)}
        \int d^dy~
        \frac
        {\partial^2 B(x)}
        {\partial (\partial^z_\rho A^{\sigma}(z)) \partial (\partial^y_{\rho^\prime} A^{\sigma^\prime}(y))}
        \partial_\mu^y \big(\partial^y_{\rho^\prime} A^{\sigma^\prime}_z(y)\big). 
\end{equation}

\section{Examples of Functional Formalism}
\label{ExamplesOfFunctionalFormalism}
In this appendix, we shall provide simple examples of the functional formalism introduced in \cref{FunctionalFormalismSection} to help illustrate how it works in practice. 
\subsection*{Example 1}
Define $F[f(\circ)]$ and $G[f(\circ)]$ as follows:
\begin{subequations}
    \begin{align}
        F[f(\circ)]
        &\equiv
        \int d^d x~
        f(x)\\
        \label{GDef}
        G[f(\circ)](x)
        &\equiv
        f^2(x).
    \end{align}
\end{subequations}
Clearly, $F$ and $G$ are examples of a functional and a function-valued functional, respectively. We may freely compose $F$ and $G$ to obtain
\begin{equation}
    \label{compositionExampleEq1}
        F[G[f(\cdot)]\circ]=F[f^2(\circ)]=\int d^dx f^2(x).
\end{equation}
Now, as illustration of the chain rule for functional derivatives, \cref{functionalDerivativeChainRule}, we shall compute the functional derivative of $F[G[f(\cdot)]\circ]$ with respect to $f(y)$ by directly using \cref{compositionExampleEq1}, and then by using the chain rule for functional derivatives in \cref{functionalDerivativeChainRule}, showing that both methods yield the same result. First, we compute the functional derivative directly from \cref{compositionExampleEq1}:
\begin{subequations}
    \label{LHSExample1}
        \begin{align}
            \label{directFunctionalDerivativeExample1}
            \frac{\partial F[G[f(\cdot)]\circ]}{\partial f(y)}
            &=
            \frac{\partial}{\partial \alpha}
            \left(
                F[G[f(\cdot)+\alpha \delta^d(\cdot-y)]\circ]
            \right)
            \bigg|_{\alpha=0}\\
             &=
            \frac{\partial}{\partial \alpha}
            \left(
                F\big[\big(f(\circ)+\alpha \delta^d(\circ-y)\big)^2\big]
            \right)
            \bigg|_{\alpha=0}\\
            &=
            \frac{\partial}{\partial \alpha}
            \left(
                \int d^dx \big(f(x)+\alpha \delta^d(x-y)\big)^2
            \right)
            \bigg|_{\alpha=0}\\
            \label{deltaSquaredTerm}
            &=
            \frac{\partial}{\partial \alpha}
            \left(
                \int d^dx \bigg\{f^2(x)+2\alpha f(x) \delta^d(x-y) +\left(\alpha\delta^d(x-y)\right)^2\bigg\}
            \right)
            \bigg|_{\alpha=0}\\
            &=
            2
            \int d^dx f(x)\delta^d(x-y)\\
            &=
            2f(y),
        \end{align}
    \end{subequations}
in \cref{directFunctionalDerivativeExample1}, we have used \cref{functionalDerivativeShorthand}. Note that in \cref{deltaSquaredTerm}, the term proportional to $(\delta^d(x-y))^2$ should be written as $\lim_{\beta \to 0}(\delta_\beta^d(x-y))^2$, where $\delta_\beta^d(x-y)$ is a regularized delta function that approaches the Dirac delta function as $\beta\to 0$. This term is in general ill-defined unless one takes the limit to $\alpha=0$ first. Now, we shall compute the same functional derivative using the left-hand side of the chain rule for functional derivatives in \cref{functionalDerivativeChainRule}:
\begin{subequations}
    \label{RHSExample1}
    \begin{align}
        \label{firstStepOfFunctionalDerivativeChainRuleExample1}
        \frac{\partial F[G[f(\cdot)]\circ]}{\partial f(y)}
        &=
        \int d^d z~
        \frac{\partial F[G[f(\cdot)]\circ]}{\partial G[f(\cdot)](z)}
        ~\frac{\partial G[f(\cdot)](z)}{\partial f(y)}\\
        &=
        \int d^d z~
        \frac{\partial}{\partial \alpha}
        \left(
            F[G[f(\cdot)]\circ+\alpha \delta^d(\circ-z)]
        \right)
        \bigg|_{\alpha=0}
        ~
        \frac{\partial}{\partial \beta}
        \left(
            G[f(\cdot)+\beta \delta^d(\cdot-y)](z)
        \right)
        \bigg|_{\beta=0}\\
        &=
        \int d^d z~
        \frac{\partial}{\partial \alpha}
        \left(
            \int d^dx f^2(x) + \alpha \delta^d(x-z)
        \right)
        \bigg|_{\alpha=0}
        ~
        \frac{\partial}{\partial \beta}
        \left(
            (f(z)+\beta \delta^d(z-y))^2
        \right)
        \bigg|_{\beta=0}\\
        &=
        \int d^d z d^dx~
        \left(
        \delta^d(x-z)
        \right)
        \left(
        2f(z)
        \delta^d(z-y)
        \right)\\
        &=
        2f(y),
    \end{align}
\end{subequations}
where in \cref{firstStepOfFunctionalDerivativeChainRuleExample1} we have used \cref{functionalDerivativeShorthand2}. We see that both methods of computing the functional derivative yield the same result, thus illustrating how the chain rule for functional derivatives works in practice.

\subsection*{Example 2}
In this example, we shall illustrate the use of the derivative chain rule in \cref{derivativeChainRule}. We define the following function-valued functional
\begin{equation}
    H[f(\circ)](x)\equiv \ln(f(x)).
\end{equation}
We may then form the composition of $G$ from \cref{GDef} and $H$ to obtain
\begin{equation}
    H[G[f(\circ)]](x)=H[f^2(\circ)](x)=2\ln(f(x)).
\end{equation}
Computing the left-hand side of the derivative chain rule in \cref{derivativeChainRule}, we find:
\begin{subequations}
    \begin{align}
        \partial_\nu^x H[G[f(\circ)]](x)
        &=
        \partial_\nu^x
        \big(
            2\ln(f(x))
        \big)\\
        &=
        \frac{2}{f(x)}
        ~\partial_\nu^x f(x).
    \end{align}
\end{subequations}
Now, we shall compute the right-hand side of the derivative chain rule in \cref{derivativeChainRule}:
\begin{subequations}
    \begin{align}
        \label{firstStepOfDerivativeChainRuleExample2}
        \sum_{j=1}^{m}
        \int d^d z~
        \frac{\partial H[G[f(\circ)](\cdot)](x)}{\partial G[f(\circ)](z)}
        ~\partial_\nu^z G[f(\circ)](z)
        &=
        \int d^d z~
        \frac{\partial}{\partial \alpha}
        \left(
            H[G[f(\circ)](\cdot)+\alpha \delta^d(\cdot - z)](x) 
        \right)
        \bigg|_{\alpha=0}
        ~
        \partial_\nu^z
        \big(
            f^2(z)
        \big)\\
        &=
        \int d^d z~
        \frac{\partial}{\partial \alpha}
        \left(
            \ln(f^2(x) + \alpha \delta^d(x-z))
        \right)
        \bigg|_{\alpha=0}
        ~
        2f(z)
        ~\partial_\nu^z f(z)\\
        &=
        \int d^d z~
        \frac{1}{f^2(x)}
        ~\delta^d(x-z)
        ~
        2f(z)
        ~\partial_\nu^z f(z)\\
        &=
        \frac{2}{f(x)}
        ~\partial_\nu^x f(x),
    \end{align}
\end{subequations}
where in \cref{firstStepOfDerivativeChainRuleExample2} we made use of \cref{functionalDerivativeShorthand2}. We see that both sides of the derivative chain rule in \cref{derivativeChainRule} yield the same result, thus illustrating how the derivative chain rule works in practice.

\section*{References}
\bibliography{refs.bib}

\end{document}